\documentstyle{mn}
\input psfig.sty
\newif\ifAMStwofonts
\AMStwofontstrue

\def\oversim#1#2{\lower0.5pt\vbox{\baselineskip0pt \lineskip-0.5pt
     \ialign{$\mathsurround0pt #1\hfil##\hfil$\crcr#2\crcr\sim\crcr}}}

\title[Irregular OH/IR stars]
      {Bipolar outflows in OH/IR stars}

\author[A.A. Zijlstra et al.]
       {Albert A. Zijlstra$^1$\thanks{e-mail: A.Zijlstra@umist.ac.uk}
    \thanks{Offprints: Albert Zijlstra},
       J.M. Chapman$^{2,3}$,  P. te Lintel Hekkert$^2$, L. Likkel $^4$, 
\newauthor F. Comeron$^5$, R.P. Norris$^2$, F.J. Molster$^6$ 
    and R.J. Cohen$^7$\\
  $^1$     UMIST, Department of Physics, P.O. Box 88,
           Manchester M60 1QD, UK \\
  $^2$     Australia Telescope National Facility, 
           P.O. Box 76, Epping, N.S.W 2121, Australia \\ 
  $^3$     Anglo Australian Observatory, P.O. Box 296, Epping, 
           N.S.W. 2121, Australia \\
  $^4$     Department of Physics and Astronomy, 
           University of Wisconsin-Eau Claire, Eau Claire, 
           WI 54702-4004, USA \\
  $^5$     European Southern Observatory, Karl-Schwarzschildstrasse\ 2,
           D-85748 Garching, Germany \\
  $^6$     Astronomical Institute `Anton Pannekoek', University of Amsterdam,
           Kruislaan 403, NL-1098 SJ Amsterdam \\
  $^7$     Nuffield Radio Astronomy Laboratories, University of
           Manchester, Jodrell Bank, Macclesfield, Cheshire SK11 9DL, UK}

\date{Accepted --- --- ---.
      Received --- --- ---;
      in original form --- --- ---}
\pubyear{2000}

\begin{document}

\maketitle

\begin{abstract}

We investigate the development of bipolar outflows during the early
post-AGB evolution. A sample of ten OH/IR stars with irregular OH
spectra and unusually large expansion velocities is observed at high
angular resolution.  The sample includes bipolar nebulae
(e.g. OH231.8+4.2), bright post-AGB stars (HD 101584) and reflection
nebulae (e.g. Roberts 22). The IRAS colour--colour diagram separates
the sample into different types of objects. One group may contain the
immediate progenitors to the (few) extreme bipolar planetary
nebulae. Two objects show colours and chemistry very similar to the
planetary nebulae with late IR-[WC] stars. One object is a confirmed
close binary.

A model is presented consisting of an outer AGB wind which is swept up
by a faster post-AGB wind, with either the AGB or Post-AGB wind being
non-spherically symetric.  The interface of the two winds is shown to
exhibit a linear relation between velocity and distance from the star,
giving the impression of an accelerating outflow. The OH data confirms
the predicted linear velocity gradients, and also reveals torus-like,
uniformly expanding components.

All sources are discussed in detail using optical/HST images where
available.  ISO data for Roberts 22 reveal a chemical dichotomy, with
both crystalline silicates and PAHs features being present.  IRAS
16342$-$3814 shows a dense torus with mass $0.1\,\rm M_\odot$ and
density of $10^8\,\rm cm^{-3}$; HST data shows four point-like sources
located symmetrically around the nebula,  near the outer edge
of the dense torus.

Lifetimes for the bipolar OH/IR stars are shown to be in excess of
$10^4$ yr, longer than normal post-AGB timescales.  This suggests that
the toruses or disks are near-stationary. We suggest that accretion
from such a disk slows down the post-AGB evolution. Such a process
could explain the link between the long-lived bipolar nebular geometry
and the retarded star.

\end{abstract}


\section{Introduction}

Stars at the end of the Asymptotic Giant Branch (AGB) experience a
phase of high mass loss during which they expel most of the material
from their outer atmospheres and form optically-thick circumstellar
envelopes (CSEs) (e.g.\ Sch\"onberner 1989; for a review see Habing
1996).  After the cessation of the mass loss, the star heats up
rapidly and eventually begins to ionize the expanding envelope.
During the post-AGB evolution, a fast wind from the hot star sweeps up
the earlier slow AGB wind to form a planetary nebula (PN) (Kwok,
Purton \& Fitzgerald 1978).

Approximately half of all PNe show bipolar structures.  A bipolar
structure may form if the slow AGB wind possesses an 
asymmetry, which will be amplified by the post-AGB wind--wind
interaction (Icke 1988).  Hydro-dynamical models predict the eventual
formation of a dense, slowly expanding equatorial torus, with a faster
outflow weakly collimated within two polar cones (Kahn \& West 1985;
Balick, Preston \& Icke 1987; Icke 1988; Frank \& Mellema 1994). The
predicted structures compare well with the observed range of
morphologies of PNe (e.g. G\'orny et al. 1999).

This interacting-wind model predicts that the AGB wind will show
deviations from spherical symmetry.  Present evidence shows that the
outer winds of AGB stars have a high degree of symmetry.  In the OH
1612 MHz survey of te Lintel Hekkert et al.\ (1991), only 16 out of
852 OH/IR stars showed irregular line profiles. Sevenster et al.\
(1997a,b) also found almost exclusively regular profiles. The inner
wind may be less symmetric, however.  Optical interferometry (e.g. Tuthill et
al.\ 1994; Bedding et al.\ 1997) has shown elongation of the stellar
photospheres, and OH main-line masers of Mira variables have shown
axi-symmetric structures (Bowers, Johnston \& de Vegt 1989; Chapman \&
Cohen 1985; Chapman, Cohen \& Saikia 1991; Chapman et al.\ 1994).
Johnson \& Jones (1991) found significant optical and infrared
polarization in AGB and post-AGB stars.

The second prediction of the interacting-wind model is that strongly
bipolar morphologies will only develop once the star is hot enough to
exhibit a fast wind. Asymmetries during the early post-AGB evolution
should not yet be amplified, since these stars are too cool (stellar
temperature between 5000 and 15000 K) to have a hot wind.  However,
images of early post-AGB objects show a large proportion of highly
bipolar morphologies, in sharp contrast with the model prediction. In
combination with the insufficient evidence for significant asymmetries
in the AGB winds, this indicates that the standard model is
incomplete.  Sahai \& Trauger (1998) therefore propose that early in
the post-AGB evolution, a fast jet-like wind operates. An alternative
possiblity is that at the very end of the AGB, the mass loss suddenly
becomes strongly equatorially enhanced, caused by e.g. the spiral-in
of a Jupiter-like companion.

In this study we present aperture-synthesis observations of the OH
maser emission from one or more of the ground-state transitions at
1612, 1665 and 1667 MHz, for 10 likely post-AGB stars with unusually
broad and irregular spectral profiles. The principal aim of this study
is to investigate the envelope geometries of post-AGB stars and to
determine whether these can be explained using a wind-wind interaction
model.

Section 2 below describes the OH properties of post-AGB stars and the
selection of targets. Section 3 describes the observations. Section 4
discusses momentum-driven wind--wind interactions and gives a
geometric model which is used to interpret the results. Section 5
describes the results and their interpretation for individual
sources. In Section 6 we discuss the nebular and stellar parameters,
lifetimes, two proposed evolutionary sequences and the possible
effects of Post-AGB accretion on the stellar evolution. Conclusions
are summarized in Section 7.

\section{Properties of post-AGB stars}

\subsection{OH Maser emission}

An oxygen-rich CSE may exhibit one or more of the OH 18-cm maser lines
at 1612, 1665 or 1667 MHz (e.g. Elitzur 1992).  For stars with
mass-loss rates above a few $\times$ 10$^{-7}$ M$_\odot$ yr$^{-1}$,
the strongest OH maser emission occurs from the satellite line
($\Delta F = +1$) at 1612 MHz (e.g. Cohen 1989) while for stars with
lower mass-loss rates the strongest emission is usually from the OH
mainlines ($\Delta F = 0$) at 1665 and 1667 MHz.

During the post-AGB phase, the OH lines may remain observable for a
period of $\sim$ 1000 years (Sun \&\ Kwok 1987).  The OH properties of
post-AGB stars have been discussed previously by Zijlstra et al.\
(1989). The maser profiles will become irregular because of the
reduced coherence amplification lengths.  The 1667 MHz emission
becomes relatively stronger and may become dominant (Field 1985, Field
\&\ Gray 1988). The OH expansion velocity may increase due to the fact
that the AGB/post-AGB wind will become faster as the spectral type of
the star becomes earlier.  Some post-AGB stars show pronounced but
non-periodic variability in the OH maser emission and this may be a
common feature of the early post-AGB stars, e.g., OH0.9+0.3 for which
the OH flux density was observed to increase over many years at a rate
of 1$\,$Jy$\,$yr$^{-1}$ (Shepherd et al.\ 1989; Zijlstra et al.\
1989).  Strong circular polarization has been detected from the
post-AGB sources OH231.8+4.2 and IRAS 16342-3814, indicating embedded
magnetic fields (e.g. Szymczak, Cohen \&\ Richards 1998).

Table 1 is a compilation of stellar sources with unusual OH maser
properties, consistent with a possible classification as post-AGB
stars.The sources were selected to have OH maser emission with a
velocity range of at least 50 km s$^{-1}$ in one or more of the OH
lines, and a spectrum with an emission plateau and/or more than two
spectral features. We used the surveys for OH maser emission from IRAS
sources with the far-infrared colours of high mass-loss AGB stars
(Lewis, Eder \& Terzian 1985; Eder et al.\ 1988; te Lintel Hekkert et
al.\ 1991; te Lintel Hekkert \& Chapman 1996).  Table 1 however does
include a small number of massive supergiant stars which have higher
outflow velocities than OH/IR stars. Examples in this category are VY
CMa, NML Cyg and PZ Cas. The large expansion velocities in our
selection criteria are not {\it a-priori} predicted for post-AGB stars
and only a biased subset of all post-AGB stars may have been selected.
For completeness we include a few carbon-rich post-AGB sources in
Table 1 with similarly high outflow velocities, although these do not
show OH maser emission.

From the sources listed in Table 1, we selected 10 sources (indicated
by an asterisk after the source name) for aperture synthesis
observations.

\begin{table*}
\begin{minipage}{160mm}
\caption{Stellar sources with unusual OH spectra. Sources discussed in
this paper are indicated with a star.}
\begin{tabular}{lccrrlrrcc}
\hline
 IRAS      & RA \hfil & Dec \hfil &  f12 \hfil  &  f25 \hfil & Name  & 
 $\Delta V_{\rm OH}$ & $\Delta V_{\rm CO}$    & Morphology & Refs \cr
           & (B1950)  & (B1950)  &         (Jy)&      (Jy)  &       &
     \multicolumn{2}{c}{(km s$^{-1}$)}           \\
04395$+$3601 & 04 39 34.1 &  +36 01 15 &  ~471~~   & 1106~~  & CRL618       
&  -  & 400  & Bipolar      & 1,2 \\
07209$-$2540 & 07 20 55.6 &$-$25 40 17 &  9919~~   & 6651    & VY CMa       
&  ~57 & ~65 &          & 3,4 \\
07399$-$1435* & 07 39 59.2 &$-$14 35 42 &  ~~19.0   & 226~~~  & OH231.8+4.2  
& 100 & 200  & Bipolar      & 5,6,7,10 \\
08005$-$2356* & 08 00 32.5 &$-$23 56 16 &  ~~18.0   & ~~51.8  &              
& 100 &   -  & Bipolar      & 8,9,12 \\
09371$+$1212 &  09 37 11.6 & +12 12 31  & ~~~~0.27 & ~~~4.6  & Frosty Leo
& -   &  80  & Bipolar      & 38 \\
10197$-$5750* & 10 19 44.8 &$-$57 50 32 &  ~200~~   & 1092~~  & Roberts 22   
&  ~52 & ~50 &          & 11,13,14 \\
10481$-$6930 & 10 48 08.8 &$-$69 30 03 &  ~~36.0   & ~~45.7  &              
&  ~51 &   - &          & 9 \\
10491-2059   & 10 49 11.4 &$-$20 59 06 &  1110     & 460~~~  &  V Hya 
&    -  & 200 &          & 34 \\
11385$-$5517* & 11 38 33.9 &$-$55 17 49 &  ~~92.6   & 138~~~  & HD 101584    
&  ~84 & 300 & Bipolar      & 10,15,16,17 \\
13328$-$6244 & 13 32 52.6 &$-$62 44 01 &  110~~~   & 144~~~  &              
&  ~50 &   - &          & 9 \\
13442$-$6109 & 13 44 17.7 &$-$61 09 30 &  389~~~   & 285~~~  &              
&  ~50 &   - &          & 18 \\
14028$-$5836 & 14 02 50.5 &$-$58 36 23 &  ~~23.6   & ~~40.4  &              
&  ~50 &   - &          & 19 \\
15303$-$5456 & 15 30 20.6 &$-$54 56 45 &  ~~40.4   & ~~59.6  &              
&  ~50 &   - &          & 19 \\
15405$-$4945* & 15 40 32.1 &$-$49 45 54 &  ~~~2.4   & ~~26.5  &              
& 175 &    - & Bipolar      & 10,20 \\
15452$-$5459 & 15 45 17.0 &$-$54 59 42 &  ~~87.1   & ~~~243  &              
&  ~80 &   - &          & 10 \\
15509$-$5207 & 15 50 59.0 &$-$52 07 41 &  ~~59.5   & ~~93.2  &              
&  ~60 &   - &          & 19 \\
16333$-$4654 & 16 33 21.8 &$-$46 54 38 &  ~~44.3   & ~~54.9  & OH337.5-0.1  
&  ~52 &   - &          & 3,19 \\
16342$-$3814* & 16 34 17.1 &$-$38 14 18 &  ~~16.2   & ~200~~  &              
& 140  &   - & Bipolar      & 10,20,21 \\
16383$-$4626 & 16 38 19.3 &$-$46 26 33 &  ~~15.0   & ~~21.2  & OH338.5-0.2  
&  54 &    - &          & 3 \\
17079$-$3844 & 17 07 57.0 &$-$38 44 40 &  ~~32.0   & ~~19.6  &              
&  ~54 &   - &          & 9 \\
17253$-$2831* & 17 25 23.6 &$-$28 31 04 &  ~~~1.4   & ~~20.1  &              
&  ~52 &   - &          & 9 \\
17393$-$3004 & 17 39 22.4 &$-$30 04 20 &  ~256~~   & ~~~218  &              
&  ~53 &   - &          & 3 \\
17423$-$1755* & 17 42 18.8 &$-$17 55 36 &  ~~~7.1   & ~~28.3  & He3-1475     
&  ~55 &   - & Bipolar      & 10,23,24 \\
           - & 17 42 33.6 &$-$28 58 03 &           &         & 359.970-0.049
&  ~50 &   - &          & 25 \\
           - & 17 43 58.9 &$-$28 28 52 &           &         & 0.548-0.059
&  ~50 &   - &          & 25 \\
18052$-$2016 & 18 05 17.9 &$-$20 16 42 &  ~~21.4   & ~~32.5  & OH10.0-0.1   
&  ~57 &   - &          & 3 \\
18091$-$1815 & 18 09 07.6 &$-$18 15 11 &  ~~22.1   & ~~38.0  & OH12.3+0.1   
&  ~67 &   - &          & 3 \\
18246$-$1032 & 18 24 37.9 &$-$10 32 30 &  ~~~2.2   & ~~20.3  & OH20.8+0.5   
&  ~55 &   - &          & 3 \\
             & 18 33 00.0 &$-$08 04 03 &           &         & OH24.0-0.2   
&  ~64 &   - &          & 3 \\
18349$+$1023 & 18 34 57.8 &  +10 23 04 &  ~720~~   & ~319~~  & IRC+10365    
&  51 &   35 &          & 3,4 \\
18491$-$0207* & 18 49 10.7 &$-$02 07 48 &  ~~~5.5   & ~~23.2  &              
& 180 &    - &          & 10,23 \\
18585$+$0900 & 18 58 30.2 &  +09 00 43 &  ~~56.8   & ~~63.5  &              
&  ~62 &   - &          & 27 \\
19052$+$1431 & 19 05 16.8 &  +14 31 59 &  ~~~2.9   & ~~~3.0  &              
&  ~59 &   - &          & 27 \\
19114$+$0002 & 19 11 25.0 &  +00 02 18 &  ~~31.3   & ~648~~  & HD 179821    
&  ~60 & ~80 &          & 8,28,29 \\
19244$+$1115 & 19 24 26.3 &  +11 15 09 &  1346~~   & 2314~~  & IRC 10420    
&  ~60 & 104 & Bipolar      & 30,31,32 \\
19500$-$1709 & 19 50 01.5 &$-$17 09 38 &  ~~27.8   & ~165~~  & SAO 163075   
&  -   & ~85 & Bipolar      & 28,13 \\
             & 21 00 19.9 &  +36 29 45 &           &         & CRL2688      
&  -   & ~85 & Bipolar      & 33 \\
             & 20 44 34.0 &  +39 55 54 &           &         & NML Cyg
&  ~68 &   - &          & 36, 37 \\
22036$+$5306* & 22 03 40.0 &  +53 06 55 &  ~~~8.4   & ~~46.3  &              
&  ~80 &   - &          & 10,21 \\
23416$+$6130 & 23 41 39.1 &  +61 30 43 &  373~~~   & ~398~~  & PZ Cas    
&  ~53 &   - &          & 3, 35 \\
\hline
\end{tabular}

\medskip 
References: 
1)  Cernicharo et al. 1989; 
2)  Kwok \& Bignell 1984; 
3)  te Lintel Hekkert et al. 1989 (and ref. therein);
4)  Loup et al. 1993;
5)  Morris et al. 1987; 
6)  Icke \& Preston 1989;
7)  Reipurth 1987;
8)  Likkel 1989;
9)  te Lintel Hekkert et al. 1991;
10) te Lintel Hekkert \& Chapman 1996;
11) Allen, Hyland \&\ Caswell, 1980;
12) Slijkhuis, de Jong \& Hu 1991;
13) Bujarrabal \& Bachiller 1991;
14) Sahai et al. 1999a;
15) This paper;
16) Trams et al. 1990;
17) Loup et al. 1990;
18) Gaylard et al. 1989;
19) Gaylard \&\ Whitelock 1989;
20) te Lintel Hekkert et al. 1988;
21) Sahai et al. 1999b;
22) Zijlstra et al. 1989;
23) te Lintel Hekkert 1991;
24) Bobrowski et al. 1995;
25) Lindqvist et al. 1992;
26) Eder, Lewis \&\ Terzian 1988;
27) Lewis, Eder \&\ Terzian, 1985;
28) Likkel et al. 1987;
29) Zuckerman \& Dyck 1986;
30) Mutel et al. 1979;
31) Diamond, Norris \& Booth 1983;
32) Knapp \& Morris 1985;
33) Kawabe et al. 1987;
34) Knapp, Jorissen \&\ Young 1997;
35) Dickinson \&\ Chaisson 1973;
36) Wilson \&\ Barret 1968;
37) Diamond, Norris \& Booth 1984;
38) Sahai et al. 2000
%
\end{minipage}
\end{table*}

\subsection{IRAS colours}

Fig.\ 1 shows the IRAS colour--colour diagram for the sources of Table
1 for which good IRAS data is available. The axes are defined as $R21
= \log{F_{25}/F_{12}}$ and $R32 = \log{F_{60}/F_{25}}$.

\begin{figure*}
\psfig{figure=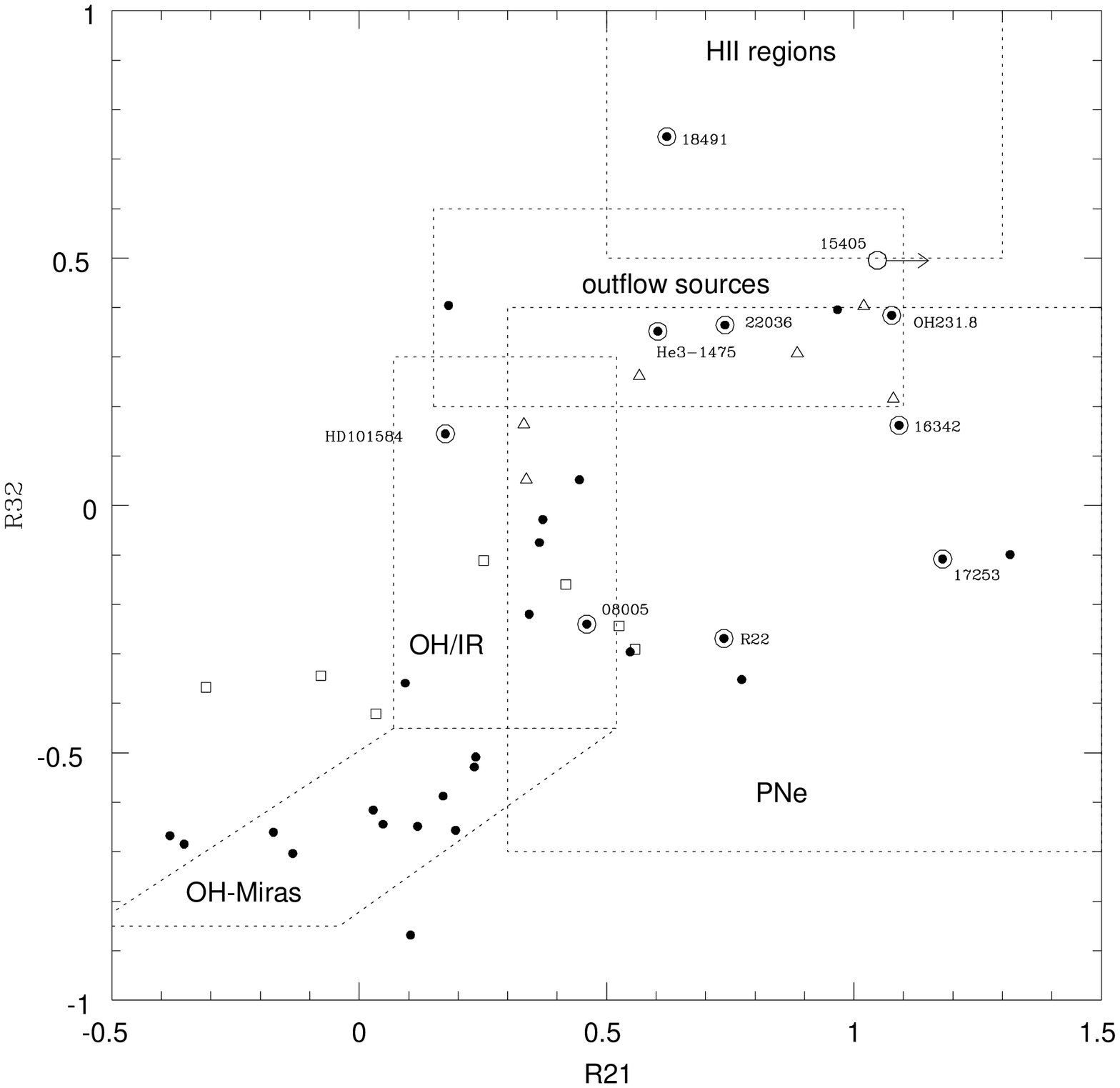,width=160mm,height=160mm}
\caption{IRAS colour--colour plot.  R21 = $\log{F_{25}/F_{12}}$
and R32 = $\log{F_{60}/F_{25}}$, with the IRAS fluxes not
colour-corrected. Regions for different types of objects are taken
from Pottasch et al.\ (1988), Zijlstra (1991) and Emerson (1987).
Filled circles are objects from Table 1. Objects discussed in this
paper are encircled. The open triangles show the extreme bipolar
planetary nebulae, NGC6302, M2-9, Hb4, He2-437, 19W32, K3-35. The open
squares show the [WC] stars with mixed C/O chemistry: BD+30$^{\rm
o}$3639, IRAS07027$-$7932, He2-113, CPD$-$56$^{\rm o}$8032 (Waters et
al.\ 1998b; Zijlstra et al.\ 1991; Cohen et al.\ 1998)
}
\end{figure*}

The sources selected for maser observations are shown as encircled
points. Some additional objects of interest, discussed in Section 6,
are also plotted. Triangles indicate highly bipolar PNe and squares
post-AGB stars with mixed C/O chemistry, which may be evolutionary
related to the present sample (see Section 6.3).  The boxes indicate
the regions of the colour-colour diagram where different types of
objects are clustered.  The box labeled 'outflow sources' contains a
mixture of both young (Emerson 1987) and evolved (e.g. Zijlstra 1991)
stars with fast bipolar outflows.

Of the ten sources discussed in this paper, three have colours
consistent with AGB/post-AGB evolution: IRAS 08005$-$2356, Roberts 22,
and IRAS 17253$-$2831. The last object has colours corresponding to
extremely high, on-going mass loss (evolved PNe may show similar
colours but are much fainter and show no OH). The colours of Roberts
22 and IRAS 08005$-$2356 can be explained in terms of a detached CSE.
One source, IRAS 18491$-$0207, has far-infrared colours consistent
with HII regions.

He3-1475, IRAS 22036+5306, OH231.8+4.2 and IRAS 16342-3814, are
located among the outflow sources. HD 101584 could also be in this
group: its colours are at best marginally consistent with normal OH/IR
stars. IRAS 15405$-$4945 was not detected at 12$\mu$m, but may also
fall among the outflow sources.

Half the selected stars have colours different from those expected for
normal post-AGB evolution.

\section{Observations}

For each of the ten sources indicated in Table 1, observations were
taken between 1986 and 1992 in one or more of the OH transitions at
1612, 1665 or 1667 MHz, using the Very Large Array (VLA), the Jodrell
Bank MERLIN array, the Australia Telescope Compact Array (ATCA) and
the Australian Parkes--Tidbinbilla Interferometer (PTI).  The
observations are summarised in Table 2, the columns of which are as
follows: \\

\noindent i) IRAS name \\
ii) OH transition observed \\
iii) rms level in the OH images (mJy)\\
iv) V$_{\rm{res}}$, the velocity resolution (km s$^{-1}$)\\
v) $\Delta$BW, the bandwidth of the spectral channels (kHz) \\
vi) $N_{chan}$, the total number of channels used \\
vii) the array used \\
viii) the longest baseline of the array (km) \\
ix) the angular resolution (arcsec) \\ 
x) the polarisation(s) observed \\ 
xi) the date of the observations \\

The VLA observations were taken using 27 25-m telescopes in either the
A or A/B array configuration, with a maximum baseline of 36 km
corresponding to a highest angular resolution of $\sim$ one
arcsecond. For all southern sources the angular resolution is lower in
the North--South direction. The observations were taken in both left
and right-circular polarisation with integration times of 10--25
minutes per source, making use of the good instantaneous u-v coverage
provided by the VLA. Several different spectral-line observing
configurations were used, with total bandwidths between 0.2 and 3.1
MHz and velocity resolutions between 0.5 and 8.8 km s$^{-1}$.
IRAS16342$-$3814 was observed at two settings for the central
velocity, because it shows two groups of maser lines with a large
separation in velocity, indicated in Table 2 as L (low) and H (high)
respectively.

MERLIN observations were taken for the source IRAS$\,$08005$-$2356
using four telescopes at Lovell, Darnhall, Knockin and Defford. The
source was observed using left circular polarisation for a total
integration time of approximately six hours. The longest baseline of
127 km, from Lovell to Defford, corresponds to a smallest angular
resolution of 0.3 arcsec. The MERLIN correlator gave 160 delay
channels per baseline.  The data were weighted with a triangular
weighting function and were Fourier transformed and averaged to give
visibility amplitudes and phases for 64 frequency channels across the
spectrum, with a velocity resolution of 0.9 km s$^{-1}$.

The ATCA data were taken using six 22-m telescopes on an East--West track 
with a maximum baseline of 6 km. Each source was observed for between 5
and 10 hours using two linear polarisations. A spectral bandwidth of 8
MHz was used, split into either 512 or 1024 spectral channels to give
a velocity resolution of either 1.7 or 3.4 km s$^{-1}$.  

The VLA, MERLIN and ATCA data were analysed using standard routines
from the AIPS radio astronomy package. The visibility amplitudes and
phases were first corrected for atmospheric amplitude and phase
variations, either by phase referencing to a nearby continuum source,
or by self calibration using the spectral channel with the strongest
OH emission. The visibility data were also corrected for bandpass
variations and any continuum emission was removed. For the VLA and
ATCA data, total-intensity (Stokes I) spectral-line cubes showing the
OH emission in each spectral channel were then obtained by
Fourier-inverting the data and using the standard clean procedure. For
the MERLIN data, images were obtained for left circular polarisation
only. For each source and OH line, the spectral line cube was searched
for emission features above a detection threshold of
three times the rms level. The positions and total flux
densities of the emission features were obtained using the AIPS task
JMFIT which is a two-dimensional Gaussian fitting procedure.

The PTI (Parkes-Tidbinbilla Interferometer) is a two-element real-time
radio-linked interferometer using the 64-m Parkes radiotelescope and
the 70-m NASA Deep Space network antenna at Tidbinbilla (Norris et
al. 1988). The resolution is 0.13 arcsec at 1612 MHz. PTI observations
of IRAS 10197-5750 (Roberts 22) were taken at 1665 MHz in 1986
December and at 1612 MHz in 1987 February, using a bandwidth of 0.5
MHz divided into 256 complex frequency channels. Left circular
polarisation was used. The observations were taken by alternating
between two sources for a period of about 12 hours, giving for each
source a total of about five hours integration time with a full
12-hour u-v coverage.  The single, long baseline of 275 km resolves
out the extended OH flux and the fringes detect the compact maser
spots. It is therefore ideal for the determining very accurate
positions for small components.

The PTI data were analysed using a technique in which one `reference'
maser feature was chosen for its strong and constant amplitude, which
was assumed to be unresolved. The phase of this reference feature was
then subtracted from the phases of all other features, thereby
performing a phase correction for all atmospheric, ionospheric, and
instrumental effects. The residual phases are a sinusoidal function of
the positional offset from the reference feature, and so the relative
positions in both right ascension and declination were obtained by
fitting a sinusoid to the phase. The typical positional accuracy is
about 0.01 arcsec in both right ascension and declination.

\begin{table*}
\begin{minipage}{160mm}
\caption{Observed sources}
\begin{tabular}{llclcrlccccc}
\hline
IRAS    &  OH	& rms  & $V_{\rm{res}}$ & $\Delta$BW 
& $N_{\rm chan}$
 & array  & longest &  angular & pol. & date \\
& line &  &  &  &  & used  & baseline & resol. & 
&      \\
&     (MHz)  & (mJy)&(km s$^{-1}$)&	(kHz)& &     &(km)     
 & (arcsec) &
&      \\
07399$-$1435    & 1667	& ~7	& 2.2   & 12.2	
& 128	& VLA	& ~30	& 1.0	& I     & 31/10/88 \\

08005$-$2356    & 1612	& ~8	& 0.9	& ~4.9	
& ~~64	& MERLIN& 128	& 0.3	& L	& 24/04/88 \\
                & 1612	& ~7	& 2.2	& 12.2	
& ~~64	& VLA	& ~30	& 1.0	& I	& 15/05/90 \\
                & 1667	& ~7	& 2.2	& 12.2	
& ~~64	& VLA	& ~30	& 1.0	& I	& 15/05/90 \\

10197$-$5750    & 1612  & ~5   & 0.4   & ~2.0 
& ~256  & PTI   & 275   & 0.1   & L    & 02/02/87 \\ 
                & 1665  & ~5   & 0.4   & ~2.0 
& ~256  & PTI   & 275   & 0.1   & L    & 07/12/86 \\               
                & 1665	& 15	& 1.7	& ~7.8	
& 1024	& ATCA	& ~~6	& 6.0	& I	& 29/06/92 \\
                & 1667	& 15	& 1.7	& ~7.8	
& 1024	& ATCA	& ~~6	& 6.0	& I	& 29/06/92 \\

11385$-$5517	& 1667	& ~5	& 3.4	& 15.6	
& ~512	& ATCA	& ~~6	& 6.0	& I	& 26/08/91 \\

15405$-$4945    & 1667	& 20	& 3.4	& 15.6	
& ~512	& ATCA	& ~~6	& 6.0	& I	& 26/08/91 \\
                & 1665  & 20    & 3.4   & 15.6 
& ~512  & ATCA  & ~~6   & 6.0   & I     & 26/08/91 \\

16342$-$3814	& 1612L	& 20	& 0.5	& ~3.0	
& ~~64	& VLA	& ~30	& 1.0	& I	& 15/05/90 \\
                & 1612H	& 25	& 0.5	& ~3.0	
& ~~64	& VLA	& ~30	& 1.0	& I	& 15/05/90 \\

17253$-$2831	& 1612	& 10	& 2.2	& 12.2	
& ~128	& VLA	& ~30	& 1.0	& I	& 30/10/88 \\

17423$-$1755	& 1667	& 20	& 1.1	& ~6.1	
& ~128	& VLA	& ~30	& 1.0	& I	& 31/10/88 \\

18491$-$0207    & 1667  & ~6	& 8.8	& 48.8	
& ~~64	& VLA	& ~30	& 1.0	& I	& 31/10/88 \\

22036+5306	& 1612	& 20	& 2.2	& 12.2	
& ~~64	& VLA	& ~30	& 1.0	& I	& 15/05/90 \\
                &  1665	& 35	& 2.2	& 12.2	
& ~~64	& VLA	& ~30	& 1.0	& I	& 15/05/90 \\
                & 1667  & 20	& 2.2	& 12.2
& ~~64	& VLA	& ~30	& 1.0	& I	& 15/05/90 \\
\hline
 \end{tabular}
\end{minipage}
\end{table*}

\section{Wind--wind models and velocity structures}

\subsection{A wind--wind model for bipolar emission}

The standard interacting wind model assumes that the fast wind has
a velocity of order $10^4\,\rm km\,s^{-1}$. However, the stellar wind
during the early post-AGB evolution is very much slower. Here we will
discuss two interacting winds with velocities up to a few 100 km/s. At
these speeds the interface between the winds is momentum-driven rather
than energy-driven.

Shu et al.\ (1991) have calculated the velocity structures expected
from a $\sim 100 \,\rm km\,s^{-1}$, spherically symmetric stellar wind
expanding into a dense medium that has a density profile of $\rho\sim
z^{-2}$, with $z$ the distance along the polar axis.  Such a model can
explain the bipolar outflows seen around young stars (e.g. Barral \&\
Canto 1981; Frank et al. 1993).  (An alternative model where bipolar
structure formation is driven by a jet is described by e.g. Hatchell,
Fuller \&\ Ladd, 1999; MHD models which create bipolar structures are
discussed in Garcia-Segura et al.\ 1991).  The swept-up material forms
two expanding bubbles on either side of the denser equatorial
region. At any point along the surface of the bubbles, the velocity is
radial with respect to the central star, and is proportional to
distance along the polar axis, $z$.  If the opening angle of the cone
does not vary greatly with $z$, the observed velocity would appear to
increase linearly with radial distance $r$ and give the appearance of
a linearly accelerating outflow.

Velocities increasing with distance from the star has been observed
both from unusual OH/IR stars (as shown in this paper) and from PNe
(e.g., MyCn18: Bryce et al.\ 1997; Fleming 1: Lopez, Meaburn \&\
Palmer\ 1993). The Shu et al.\ model cannot be applied directly to
post-AGB stars or PNe. However, the {\it observed} similarities
between the two categories suggest that a similar model may apply to
post-AGB stars: Jet-like structures observed in a few post-AGB stars
bear a striking resemblance to Herbig-Haro objects associated with
young stars (Bobrowski et al.\ 1995; Riera et al.\ 1995; Bujarrabal et
al.\ 1998), and their IRAS colours (Fig. 1) are also very similar.  To
adapt the Shu et al.\ model to the post-AGB evolution, we discuss the
problem where a faster, inner wind (subindexed $w$) collides with a
slower, outer wind (subindexed $AGB$), at velocities sufficiently low
that the wind--wind interaction is momentum-driven. Both the AGB wind
and the faster inner wind are assumed to be expanding at terminal
(time-independent) velocity, $V$ and mass-loss rate, $\dot M$, where
$V$ and $\dot M$ of either wind may have a polar dependence.  The
wind--wind collision produces a compressed shell of material which
extends into the outer wind. To allow for the polar dependence, we
define an {\it effective mass-loss rate} ${\dot m}$:

$$ {\dot m(\theta)} = {\dot M} f(\theta)\,; \qquad \int f(\theta) d
\Omega = 4 \pi, $$

\noindent where $ d \Omega$ is the element solid angle centred on the
star and $\theta$ is the angle with the pole. The density $\rho$ of the
wind at a distance $r$ from the star is

  $$\rho(r, \theta) = {{\dot m(\theta)} \over {4 \pi r^2 V(\theta)}}.
\eqno (1)$$

The compressed shell moves at a given moment with a velocity
$v(\theta)$, and has a surface density $\sigma(\theta)$. It is
composed of gas from the outer wind, swept up at a velocity $v -
V_{AGB}$ relative to the shell, and gas from the inner wind which
reaches the shell with a velocity $V_w - v$.  All the gas flows are
assumed to be radial, so that each radial direction, characterized by
$\theta$, can be treated independently. 
Mass conservation 
implies

  $${{d} \over {d t}} (4 \pi \sigma r^2) = 
   4 \pi r^2 \rho_{AGB} (v - V_{AGB}) + 4 \pi r^2 \rho_w (V_w - v) \eqno (2)$$

\noindent and  momentum conservation implies that 

$$ {{d} \over {d t}} (4 \pi \sigma r^2 v) = 
   4 \pi r^2 \rho_{AGB} V_{AGB} (v - V_{AGB})
 + 4 \pi r^2 \rho_w V_w (V_w -v), 
     \eqno (3)
$$
 
\noindent where all variables implicitedly depend on $\theta$.  Developing the
derivatives and using eq. (1), we obtain:

  $$8 \pi \sigma r v + 4 \pi r^2 {{d \sigma} \over {d t}} =
    {{{\dot m}_{AGB}} \over {V_{AGB}}} (v - V_{AGB}) + 
    {{{\dot m}_w} \over {V_w}} (V_w - v)  \eqno (4)$$

  $$4 \pi r^2 v {{d \sigma} \over {d t}} + 8 \pi \sigma v^2 r +
    4 \pi \sigma r^2 {{d v} \over {d t}} =
    {\dot m}_{AGB} (v - V_{AGB})
 +  {\dot m}_w (V_w - v).  \eqno (5)$$

  Multiplying eq. (4) by $v$, isolating the term with the derivative of 
$\sigma$ in eq. (4), and replacing it in eq. (5), yields:

$$
 {{{\dot m}_{AGB}} \over {V_{AGB}}} v (v - V_{AGB}) + 
    {{{\dot m}_w} \over {V_w}} v (V_w - v) +
    4 \pi \sigma r^2 {{d v} \over {d t}}  
$$
$$
 = {\dot m}_{AGB} (v - V_{AGB}) + {\dot m}_w (V_w - v).  \eqno(6)
$$

\noindent  We now have the system

  $$4 \pi \sigma r^2 {{d v} \over {d t}} =
    v^2 \left({{{\dot m}_w} \over {V_w}} - 
    {{{\dot m}_{AGB}} \over {V_{AGB}}}\right) + 
    2 v ({\dot m}_{AGB} - {\dot m}_w) 
 $$
$$
   + ({\dot m}_w V_w - {\dot m}_{AGB} V_{AGB})  \eqno (7)$$
    
  $$4 \pi r^2 {{d \sigma} \over {d t}} =
    v \left({{{\dot m}_{AGB}} \over {V_{AGB}}} -
    {{{\dot m}_w} \over {V_w}} - 8 \pi \sigma r\right) +
    ({\dot m}_w - {\dot m}_{AGB}).  \eqno (8)$$

In general, $r$, $v$, $\sigma$ of the compressed shell are functions
of time. A power-law dependence with time often provides a
well-behaved solution to the evolution of radius, velocity, and
surface density in scale-free scenarios involving a central source of
momentum and energy, as is the present case. Therefore, we have tried
to find a non-trivial solution by assuming that power law solutions
are possible. If power laws are solutions of eqs. (7) and (8), then
both the left- and right-hand sides of eqs. (7) and (8) must also be
power laws. Given that the right-hand sides of these equations contain
constant terms as summands, the only way to satisfy this condition is
that the products of power laws on the left-hand sides be independent
of time. Now consider the right-hand side of eq. (7), where the only
quantity that may vary is the expansion velocity $v$: the only way of
producing a time-independent combination of the quantities that appear
there is by assuming that $v$ is time-independent, which in turn
implies that both sides of eq.\ (7) are zero. Taking this into
account, the right-hand side of eq. (8) becomes time-independent if
$\sigma \propto t^{-1}$, so that the product $\sigma r$ is
constant. If this is the case, then the product $r^2 d \sigma / d t$
that appears on the left-hand side of eq.  (8) is indeed constant and
non-zero, proving that the proposed solution is indeed
consistent. 

For a time-independent $v$, the right-hand side of eq. (7) is
zero. This provides a second-degree equation in $v$, whose meaningful
solution is

  $$v = V_{AGB} {1 \over {1 - \mu / \xi}}
  \left[ 1 - \mu + (\xi - 1) \sqrt{\mu / \xi} \right],  \eqno (9)$$ 
 
\noindent where 
 
  $$\mu = {{{\dot m}_w} \over {{\dot m}_{AGB}}} \quad ; \qquad
  \xi = {{V_w} \over {V_{AGB}}}.  \eqno (10)$$

All the terms in eq. (9) may be direction-dependent. 
The other variables of the problem can be evaluated in a straightforward
manner:
  
  $$ r(\theta,t) = v(\theta)\, t  \eqno (11)$$ 

  $$\sigma ={{{\dot m_{AGB}} (v/V_{AGB} - 1) - {\dot m}_w(v/V_w-1)} \over
  {4 \pi v^2}} {1 \over t}.   \eqno (12)$$

A time-independent $v$ implies that $r \propto t$ since $v = dr /
dt$. A polar dependence in one of the two winds will lead to a
non-spherical compressed shell, with $r$ a function of $\theta$.  {\it
We therefore find that the velocity $v(\theta) \propto r(\theta)$}. The
linear velocity gradients found by Shu et al. for young objects are
also recovered for evolved stars. Below we will use the term `linear
outflow' for such a component. Note that the relation does not assume
any particular geometry, nor does it depend on which wind has the polar
dependence.

The velocity of the compressed shell $v$ can have large variations
with $\theta$, even if $\mu$ is small (as is expected for a post-AGB
wind) and $V_{AGB}$ is constant, since the term $(\xi - 1) \sqrt{\mu /
\xi}$ in the square brackets can have an amplitude of order unity.
Eq. (10) shows that $v$ will increase significantly over $V_{AGB}$ if,
in some direction $\theta$, the momentum in the fast wind becomes
comparable to the momentum in the AGB wind: $\mu \xi \sim 1$.  As an
example, if $\mu=0.1$ and $\xi=10$, then $v=1.8 V_{AGB}$; for $\xi=2$,
$v=1.1 V_{AGB}$.  This estimate shows that $v$ (and therefore $r$) may
vary by a factor of the order of 2 over the shell.  Much larger values
would require a much faster post-AGB wind for which energy-driven
models such as Icke (1988) are more appropriate.

As an aside, the significant variations possible in $v$ are a
consequence of having maintained the term ${\dot m}_w V_w - {\dot
m}_{AGB} V_{AGB}$ in eq. (7); if this term is zero or negligibly small
in {\it all} directions, and $V_{AGB}$ is independent of $\theta$,
then it is possible for the two winds to have a large density
dependence on $\theta$, but for the compressed shell to be nearly
spherical. In other words, if the two winds are self-similar in their
density distributions, then a spherical shell can result even when
the density distributions are highly direction-dependent.

\subsection{The model applied to OH/IR stars}

The model above calculates the true space velocity and true distance
from the star for each point on the compressed shell. But only the
projected radial velocity and the projected distance can be
observed. Therefore a geometrical model is required to obtain
observable quantities, even though the calculations above are valid
for any geometry.

For `classical' OH/IR stars, the OH 1612 MHz maser distributions are
located in a thin shell of constant radius, $R_{\rm shell}$, expanding
at a constant velocity $V_{\rm exp}$ (e.g. Booth et al.\ 1981; Diamond
et al.\ 1985; Chapman \& Cohen 1985; Welty, Fix \& Mutel 1987).  (The
physical location of this shell is determined by the radius at which
the external interstellar UV radiation field causes dissociation of
H$_2$O, which is the source of the OH (Huggins \&\ Glassgold 1982),
and by the internal infrared radiation field, expecially the 35-$\mu$m
photons which pump the maser inversion.)  At the extreme blue and
red-shifted velocities the strongest maser emission occurs from small
regions at the front and back of the envelope centred on the
line-of-sight through the star, where the amplification path length is
longest. At an intermediate velocity, $V$, weaker emission occurs from
a ring of projected radius $R$ given by

$$R = R_{\rm shell}\left[1 - {(V-V_*)^{2}/V_{\rm exp}}^2\right]^{0.5}
\eqno (13) $$

\noindent where $V_*$ is the stellar velocity (e.g.\ Reid et al.\
1977; Herman 1983; Herman et al.\ 1985). Plots of $R$ against $V$ 
may be used to determine the parameters $V_*$, $V_{\rm exp}$ 
and $R_{\rm shell}$.

In the interacting-winds scenario, OH emission may be observed from
the swept-up shell (the linear outflow) or from the outer undisturbed
AGB wind. The swept-up shell will contain OH if its radius is larger
than the H$_2$O dissociation radius, and  the OH masers will travel with
the swept-up gas (in contrast to AGB stars where the OH maser shell
remains at the same radius).  If the star is in the post-AGB phase,
there are two other differences with AGB shells: (1) the present,
faster wind may be atomic rather than molecular; (2) the hotter star
may dissociate the H$_2$O molecules also from the inside leaving the
entire shell OH-rich. As a consequence, the OH maser shell in the
outer AGB wind will now also expand with the gas. 

\begin{figure*}
\psfig{figure=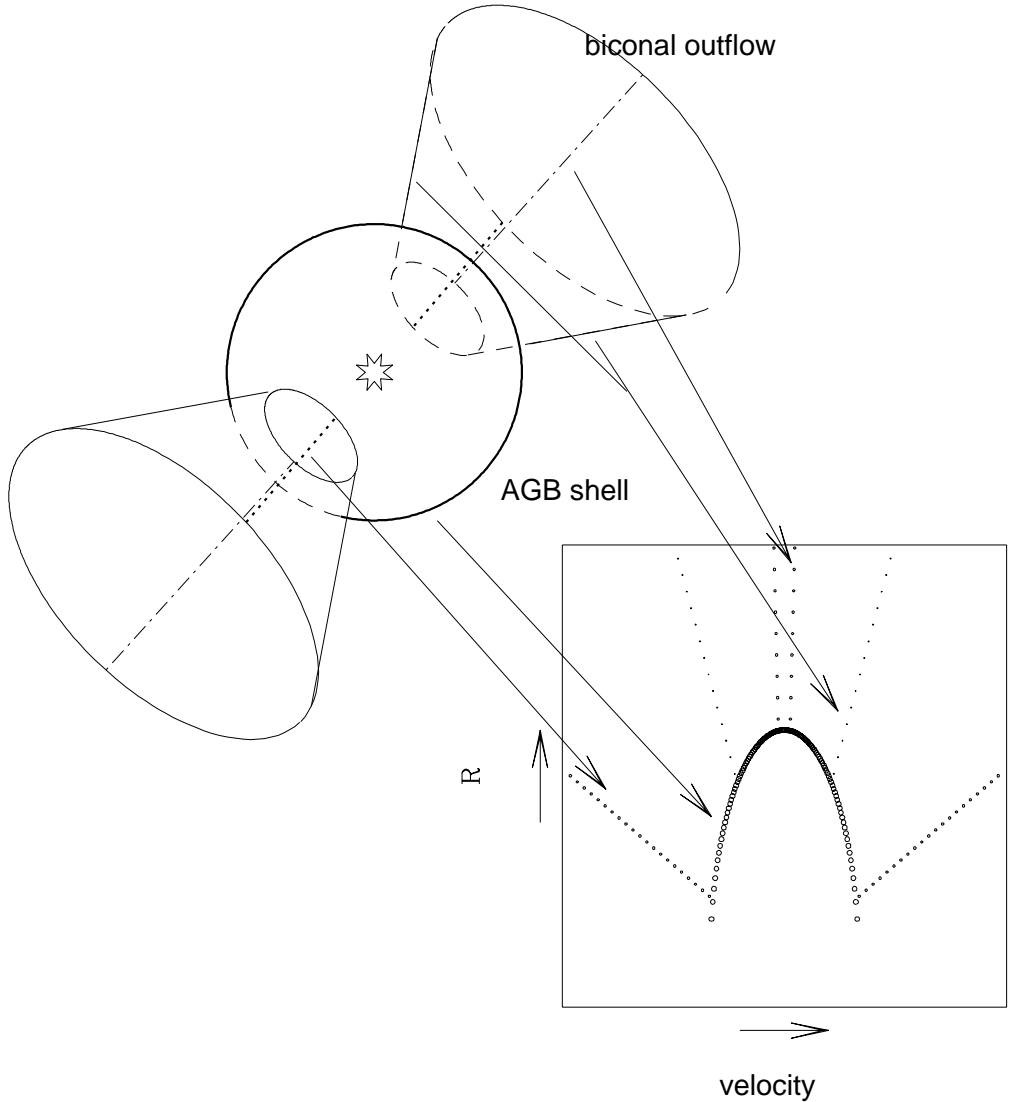,width=160mm,height=160mm}
\caption{(a)  Predicted velocity--radius relations for the model of
an hourglass-shaped bipolar outflow.
The geometric model
consisting of an AGB shell and a biconal, symmetric
outflow. Velocities are proportional with distance from the star at
each point, and are radial with respect to the star.  The dotted lines
show the location of the strongest radially-beamed emission with the
highest observed velocity gradient; the dot-dashed lines show the opposite side
of each with lowest observed velocity gradient. The separate panel shows a
velocity-radius diagram with the arrows indicating the contributions
from the different locations in the model. The elliptical distribution of points 
corresponds to the AGB shell.
}
\end{figure*}
\setcounter{figure}{1}
\begin{figure*}
\psfig{figure=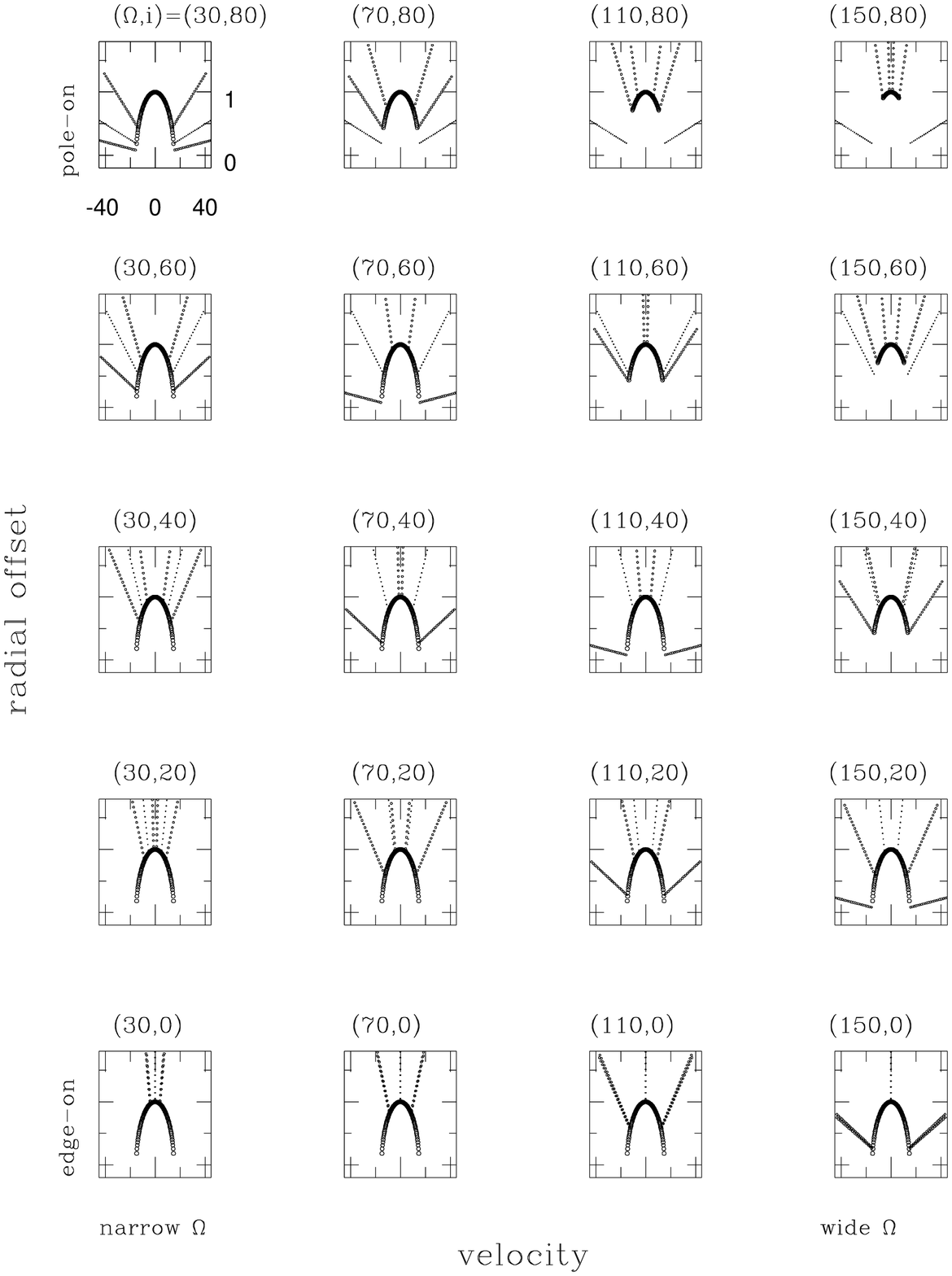,width=160mm,height=200mm}
\caption{
(b) Predicted  velocity--radius relations for various values
of $\Omega$ (the opening angle of the cone) and $i$ (the inclination
angle between the equatorial plane and the line of sight). 
See discussion in the text.}
\end{figure*}

The assumed geometry is indicated in Fig. 2a, and consists of a
spherically symmetric, thin shell representing the outer AGB wind
(expanding uniformly in all directions), with superposed a biconal or
`hourglass'-shaped structure (the wind bubbles) representing the
interaction between the two winds.  The AGB shell (which may also
represent a swept-up shell, especialy in the post-AGB phase) is
punctured by the cones: if the cones are wide, it becomes cylinder- or
torus-like.  At large radii the bubbles are expected to close (Icke
1988): this is not included in our model.  Along the surface of the
cones, the gas flows are radial away from the central star, with
velocity increasing linearly with distance from the star.  The linear
outflow is assumed to start at the radius of the thin torus/AGB shell,
with continuity in the velocity at that radius. The separate panel in
Fig. 2a shows the predicted $V$-$R$ relation for different parts of
this geometry\footnote{Note that $R$ is defined as the angular
distance between star and shell, whereas $r$ (Section 3) is defined as
the linear radius of the shell.}.

The AGB shell gives rise to the elliptical distribution of points in the
$V$-$R$ panel. If radial amplification dominates, the strongest OH
emission tracing the linear outflow will come from the lines on the
surface of the cones closest to the line of sight (dotted lines in
Fig. 2a). (Neutral gas is optically thin at these cm wavelengths so
that both the front and the back will be visible in OH.)  Emission
located elsewhere on the cones is less beamed and would normally be
fainter or invisible, with a few exceptions: (1) If the maser is
amplifying continuum emission from within the cone, the emission from
the facing sides of both cones will become stronger (the dotted line
on the blue-shifted cone and the dash-dotted line on the red-shifted
cone).This may occur if shock-ionization has occurred within the fast
wind, filling the cone with ionized gas. (2) If the interface is thin,
tangential emission may dominate. In this case the strongest emission
would come from the sides of the cone as seen projected on the sky.
(3) If the surface of the cones is very thin, non-beamed emission
would be seen from the entire surface of the cone. At each distance
$R$, this emission would fall within a velocity range determined by
the opening angle of the cone.

Radial amplification gives rise to the highest observed velocity
gradient (the most {\it horizontal} lines in the panel), symmetric for
both cones.  If continuum amplification occurs, the lowest velocity
gradient will be observed for the red-shifted cone. 

The precise observed $V$-$R$ relations depend on the opening angle of
the outflow cones, $\Omega$, and the inclination angle between the
line of sight and the equatorial plane, $i$.  Fig. 2b shows the
predicted diagrams for a range of ($\Omega, i$) pairs.  The projected
edges of the cones are shown using smaller symbols. If the
line-of-sight lies within the opening angle of the cone then the torus
will not be seen at the maximum outflow velocity: e.g., the top row of
diagrams with $i$ = 80$^0$ and large $\Omega$, where the line-of-sight
is close to the centre of the maser cones.

\begin{table*}
\begin{minipage}{160mm}
\caption{Component analysis for all observed OH/IR stars.
$V_\ast$ (in the local standard of rest) is the stellar velocity.
$V_{exp}$ is the expansion velocity of the
shell/torus component (half the total velocity width) and $R_0$
the radius of this component. For the linear outflow we give
the maximum distance from the star at which it is seen ('length')
and its maximum velocity with respect to the star ($V_l$).
OH line transitions refer to VLA to ATCA data unless indicated
(PTI or  Merlin).
}
\begin{tabular}{lclcccccl}
\hline
IRAS name	& common name	& line	& $V_\ast$       & $V_{exp}$
& $R_0$     & length  & $V_l$  &  {star}\\
		&		& MHz    & km s$^{-1}$   &   km s$^{-1}$   
& arcsec & arcsec & km s$^{-1}$ & class  \\	
iras07399$-$1435	& OH231.8+4.3	& 1667 & 35     &  35    
& 2     & 5.5    & 55 &  M9I/III, + B? \\
iras08005$-$2356	&		& 1612 & 50    & 40     
& 3.2     &  -   & -  & F5I  \\
		&		& 1667 & -   &    -   
&  -      & -    & - \\
&		& 1612Merlin & -     & 45    
& -     &  -  &  - \\
iras10197$-$5750	& Roberts 22	& 1665 & 3     &  20    
& 0.8    &  -  &  -  & A2I \\
		&		& 1667 & -     &  20     
& 1    & 1.5  &  35 \\
	             &           & 1612PTI & 0:    & 20   
&  0.8  & 1.1 & 35   \\
	             &           & 1665PTI &      &  20    
& -     &  -  &  - \\
iras11385$-$5517	& HD 101584	& 1667 & 40     & - &
 -  & 2.2 & 40  & B9II, binary \\
iras15405$-$4945	&	& 1665 & 60    & -   &  -
&    1     & 45 \\
		&		& 1667 &  60    &  -    &  - 
&    1       & 80 \\
iras16342$-$3814 & Waterfountain & 1612 & 50 & - & - 
& 1 & 70  & B? \\
iras17253$-$2831	&		& 1612	&   $-$62  &  9 & 0.25
&  0.15 & 16 \\
iras17423$-$1755	& He3-1475 & 1667	  &  50 &   25   &  0.5
&  -  & - & B[e] \\
iras18491$-$0207 &	&	1667	& 75 & - & - 
&  1 & 70 \\ 
iras22036+5306	&	& 1612	& $-$45 & - & - 
& - & - \\
		&	& 1665	& $-$45 & 20 & 1.0
& 0.5 & 15 \\
		&	& 1667	& $-$40 & 25 & 0.7 
& 1 & 30 \\
\hline
 \end{tabular}
\end{minipage}
\end{table*}

\subsection{Data representations}

For each source and observed OH transition, channel maps were made
showing the emission in each frequency channel, corresponding to a
radial velocity interval. Fig.\ 3 shows the OH 1667
MHz channel maps for OH231.8+4.2. The three-dimensional data set
(right ascension, declination, and velocity) is a projection of the
six-dimensional structure (the three-dimensional geometry of the
structure has been projected into the two dimensions of the sky, and
only the radial component of velocity is observed).  The structure is
only visible where the gain of the OH maser is sufficient to produce
detectable emission. 

\begin{figure*}
\psfig{figure=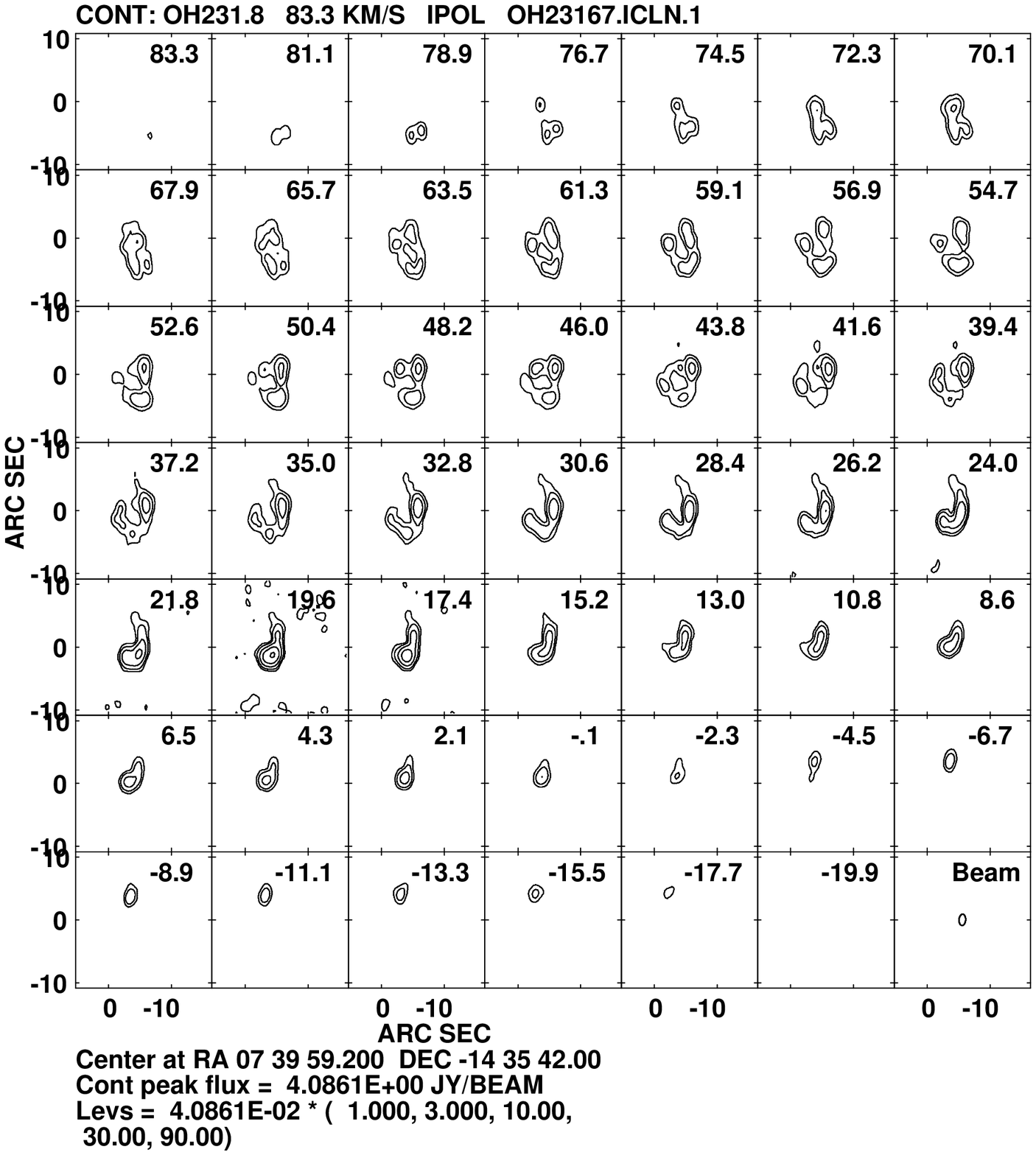,width=140mm,height=150mm}
\caption{
OH 1667 MHz channel maps of OH231.8+4.2 obtained from
VLA observations taken in 1988. The velocity of each image is
labelled in the top-right corner. The restoring beam is shown in the
last box.
}
\end{figure*}

We determined the positions of the masers at each velocity by fitting
two-dimensional Gaussian components to the detected emission in each
channel map.  This procedure recovers unresolved or slightly resolved
emission but may miss faint, more extended emission or structures such
as rings. The position of each component was typically measured to
10\% of the FWHM of the restoring beam.  We thus reduce the data to a
set of delta functions each having an associated position ($x,y$) and
velocity ($V$).  For each maser position we also determine its
projected offset from the stellar position, $R(x,y)$ where

$$ R (x,y) =  \left[ (x - x_0)^2 + (y - y_0)^2 )\right]^{0.5} \eqno (14)$$

\noindent In most cases the stellar position, ($x_0, y_0$), is
{\em not} accurately known and some assumption about the location
of the stellar position relative to the maser positions must be made
(section 5). 

\section{Individual sources}

In this section we discuss each of the 10 irregular sources for which
we have obtained aperture synthesis images. Of these, OH231.8+4.2
(IRAS 07399$-$1435) is the best resolved and is discussed in most
detail.  For each source we show `diagnostic' plots of the maser
spatial distributions ($x$-$y$), the projected separations of the masers
against velocity ($R$-$V$) and spectra showing the total integrated
emission detected in each velocity channel ($I$-$V$). The results for
all sources are summarized in Table 3. For two sources we also discuss
HST and ISO results.

\subsection{OH231.8+4.2 (IRAS 07399$-$1435)} 

The well-studied nebula OH231.8+4.2 (also known as the Rotten-Egg or
Calabash nebula) shows two ionised bipolar lobes on either side of a
central obscuring lane (Reipurth 1987). The bipolar axis is at a
position angle of 20$^{0}$. The southern, red-shifted lobe is more
extended than the northern, blue-shifted lobe, with a total extent
(both lobes) along the bipolar axis of $\sim$ 50 arcsec. Shock-excited
emission from Herbig-Haro features occurs from the front edges of both
lobes (Cohen et al.\ 1985; Reipurth 1987). The obscured central star
(QX Pup, Kastner et al. 1998) is classified as M9III or M9I (Feast et
al. 1983), and shows Mira-like variability, indicative of an evolved
AGB star. The stellar period has increased from 648 to 708 days over
20 years (Feast et al., Kastner et al., Bowers \& Morris 1984). An
excess of blue continuum emission indicates the presence of a
companion star (Cohen et al.\ 1985). Morris et al. (1987) found an
anomalously high sulphur (hence rotten egg) abundance, and suggest
that a nova explosion may have occured in the system.

The bipolar nebula of OH231.8+4.2 is a rich source of molecular
material, with a total molecular mass in the range 0.5--1 M$_\odot$
(Alcolea, Bujarrabal \& Sanch{\'e}z Contreras 1996; Sanch{\'e}z
Contreras, Bujarrabal \& Alcolea 1997). The strongest thermal
molecular emission, from $^{12}$CO, is detected at velocities between
$-75$ and 250 km s$^{-1}$ with a strong central peak at $\sim$ 35 km
s$^{-1}$ which is also seen in other molecular lines. We take this
velocity to be the stellar velocity. Approximately half of the
envelope mass is located within an unresolved ($<$ 10 arcsec) central
region at velocities within 25 km s$^{-1}$ of the stellar velocity. At
higher and lower velocities the CO emission is extended along the
bipolar axis over the full extent of the optical nebula with a
systematic outwards velocity gradient of 6 km s$^{-1}$ arcsec $^{-1}$,
with blue-shifted emission from the northern lobe and red-shifted
emission from the more extended southern lobe. For an inclination
angle of the bipolar axis to the plane of the sky of 40$^0$ (Kastner
et al.\ 1992), the maximum deprojected outflow velocity of the CO
emission, from the southern lobe, is $\sim$ 300 km s$^{-1}$.

OH231.8+4.2 was the first `irregular' OH/IR star to be discovered
(Turner 1971; Cohen et al. 1985; Cohen \&\ Frogel 1977). The OH
emission is strongest at 1667 MHz with a broad emission plateau
between $-20$ and $+80$ km s$^{-1}$ and an emission spike at $+20$ km
s$^{-1}$. From VLA images, Morris, Bowers \& Turner (1982) showed the
OH emission to be concentrated in the equatorial plane, with weak
emission also detected above and below the plane. Bowers \& Morris
(1984) found a distance of $\sim$ 1.3 kpc from OH phaselag
measurements. At this distance, the star is a likely member of the open
cluster M46 (Jura \& Morris 1985) with a progenitor mass of
approximately 3 M$_\odot$, determined from the turn-off mass of the
cluster. An initial stellar mass of at least 3 M$_\odot$ has also been
inferred from an overabundance of nitrogen in the Herbig Haro knots
(Cohen et al.\ 1985). 

Fig.\ 3 shows the individual channel maps for the 1667 MHz
observations taken with the VLA in 1988. The source was well resolved
with OH emission detected within a region of maximum north-south
extent $\sim$ 10 arcsec. Across this region there is a general
velocity gradient with the most blue-shifted masers in the north and
the most red-shifted masers in the south. At almost all velocities the
images show a well-defined symmetry axis at a position angle of
20$^{0}$, corresponding to the bipolar axis of the nebula. Near the
stellar velocity of 35 km s$^{-1}$, the OH images show ring-like or
v-shaped structures.

A comparison of the images shown in Fig.\ 3 with those of Morris et
al.\ (1982) shows no detectable changes in the maser structure between
1981 and 1987. From the earlier data, Bowers (1991) modelled
OH231.8+4.2 with an ellipsoidal density distribution, and an OH
expansion velocity of $\sim$ 100 km s$^{-1}$, several times higher than
normally seen in OH/IR stars (e.g. Baud \&\ Habing 1983). Here we
interpret our data using the two-component model discussed in section 4.

The $x$-$y$, $R$-$V$ and $I$-$V$ plots for OH231.8+4.2 are shown in
Fig.\ 4a. From the $x$-$y$ diagram it can be seen that most of the
maser centroids are located within a tilted disk or torus-like
structure. The axial ratio of the central region of $\sim$ 0.4 is
consistent with an inclination angle of $\sim 30^{\rm o}\pm10^{\rm o}$
between the line-of-sight and the plane of the disk.

\begin{figure}
\psfig{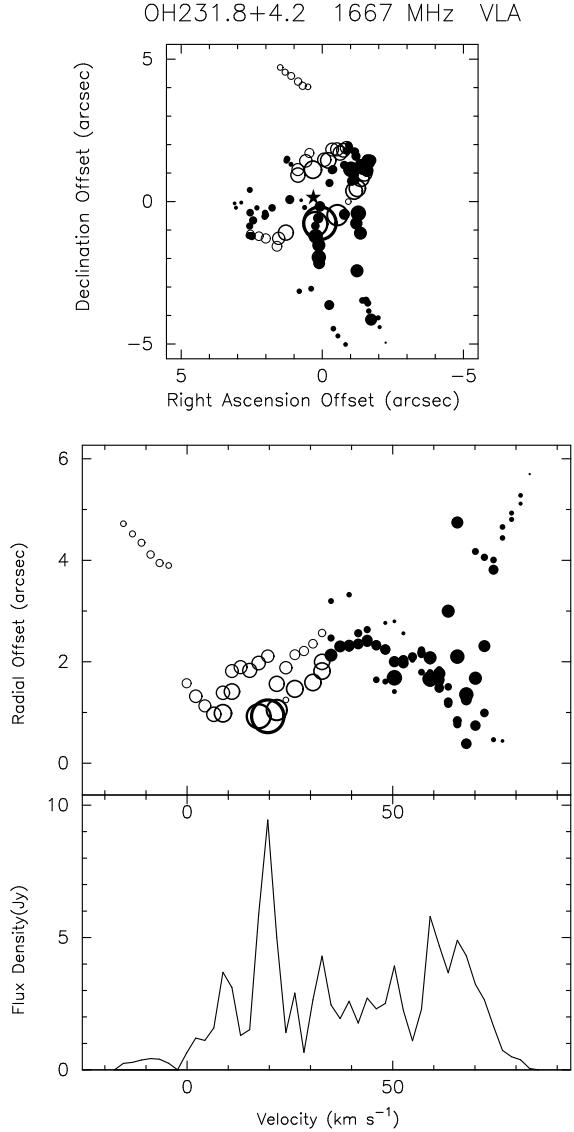}
\caption{(a) Diagnostic $x$-$y$, $R$-$V$ and $I$-$V$
diagrams for OH231.8+4.2. In this and similar plots, the top panel
shows the relative position of each emission component plotted
relative to the unweighted emission centroid. The open and filled
symbols show blue-shifted and red-shifted emission components
respectively.  The stellar position is indicated by the filled star.
The middle panel shows the projected radial offsets of the maser
components from the stellar position, plotted against observed
velocity (with respect to the local standard of rest). The bottom
panel shows the total flux density, $I$, seen in the channel maps at
each velocity.
}
\end{figure}

For OH231.8+4.2, the stellar position is not known to a high accuracy
relative to the OH masers. In the following discussion we assume that
the stellar position is located at the centre of the OH torus. From
the absolute maser positions we estimate the stellar position (J2000)
to be at: $\alpha =\rm  07^h 39^m 58.925^s$, $\delta = \rm -14^o 35^\prime 
42.4^{\prime\prime}$. This
position is indicated by the filled star in Fig.\ 4a.

{}From the $R$-$V$ diagram we identify two separate kinematic
structures corresponding to a central `torus' and a bipolar outflow,
in agreement with the models of Fig. 2. The ring or torus is expanding
outwards with a maximum expansion velocity of 35 km s$^{-1}$ and outer
radius of 2.5 arcsec. The maser velocities in this region appear
better constrained at red-shifted velocities. The scatter of points
within the toroidal shell, which is larger at blue-shifted velocities,
may indicate a velocity gradient within the torus, or possibly the
presence of multiple shells.  The precise distribution of points in
the $R$-$V$ diagram depends on the adopted stellar position but a
larger scatter in the blue-shifted maser spots is evident for any
reasonable choice of stellar position.  The maximum expansion velocity
of $\sim$ 35 km s$^{-1}$ estimated for the torus is higher than normal
for OH/IR stars also indicating that some acceleration is likely to
have occured in the torus.

Weaker blue- and red-shifted emission is seen above and below the
equatorial plane.  In the $R$-$V$ diagram these reveal a bipolar outflow
with blue- and red-shifted emission detected from the northern and
southern lobes respectively. Across these features, the OH maser
velocities increase linearly with distance from the star reaching
velocities of 50 km s$^{-1}$ (with respect to the stellar velocity),
at a radial offset of 5.5 arcsec. The OH velocity gradient of 10 km
s$^{-1}$ arcsec $^{-1}$ detected over 10 arcsec is in the same
direction but steeper than the average gradient of 6 km s$^{-1}$ seen
in CO emission over 50 arcsec.

\setcounter{figure}{3}
\begin{figure}
\psfig{figure=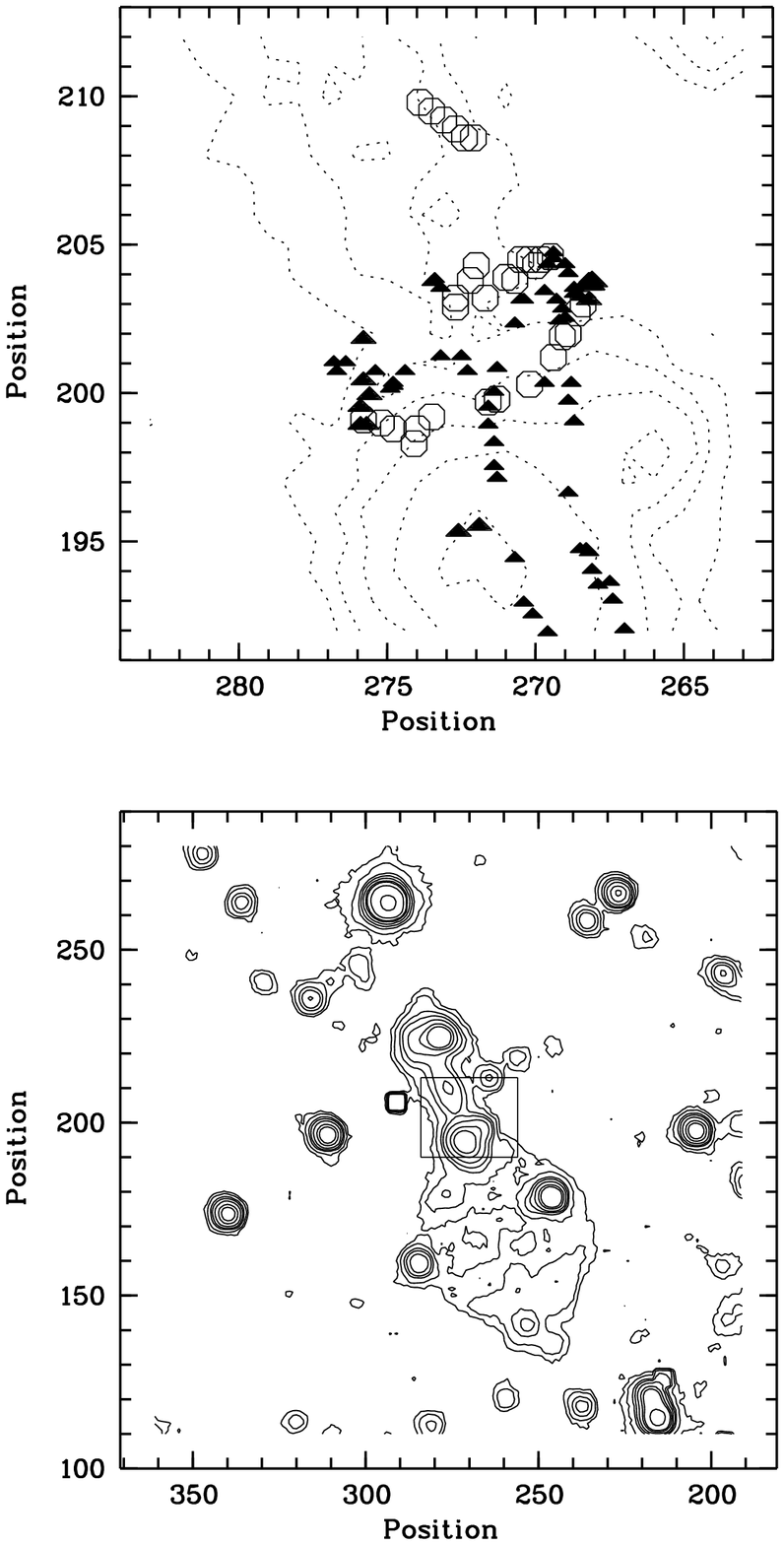,width=85mm} 
\caption{
(b) OH231.8+4.2. Bottom panel: an optical image taken through an
H$\alpha$ filter, retrieved from the La Palma archive.  The axes
labelling is in pixels and the scale is 0.55 arcsec per pixel. North is 
at the top and East to the left. 
Top panel: Overlay of the OH maser positions on the central region of the
H$\alpha$ image. The blue- and red-shifted masers are shown as open
circles and filled triangles respectively.
}
\end{figure}

Fig.\ 4b (bottom panel) shows an optical H$\alpha$ image of
OH231.8+4.2 which we retrieved from the La Palma archive.  The pixel
scale is 0.55 arcsec per pixel. The area indicated in the centre is
shown enlarged in the top panel, overlaid with the OH maser
positions. The images were aligned using stellar positions from the
Digital Sky Survey to an estimated precision of 0.3 arcsec. The OH
positions for masers within the torus agree closely with the central
dark lane discussed by Reipurth (1987).  It is likely that the CO
emission detected within $\sim$ 25 km s$^{-1}$ of the central velocity
is also associated with the central torus, with a total molecular mass
in the torus of at least $\sim$ 0.2 M$_\odot$.

In Fig.\ 4b the bipolar OH features are well aligned 
with the optical lobes but only trace the inner few arcsec of each 
lobe. A comparison of the OH positions with the models shown
in Fig.\ 2 indicates that the OH masers are detected from the front
side of the blue-shifted cone and the back-side of the red-shifted
cone where the observed velocity gradients are steepest. For a moderate cone 
opening angle (Fig.\ 2) we estimate the inclination angle of the 
OH cones to be approximately 40 degrees in good agreement with 
previous values for the inclination angle from Reipurth (1987) and 
Kastner et al.\ (1992). The opening angle of the OH cones is not well 
constrained but we suggest that the bipolar OH masers are located on
the surface of the optical lobes, near the base of the bipolar outflows.

\subsection{IRAS 08005$-$2356} 

Slijkhuis, de Jong \&\ Hu (1991) selected this source as a bright IRAS
source with far-infrared colours consistent with a detached shell, and
identified it with a $m_V=11.5$ supergiant.  Bakker et al. (1997)
assigned a spectral type of F5I and a mass-loss rate of $10^{-4.7}$
M$_\odot$ yr$^{-1}$ with an expansion velocity of 50 km s$^{-1}$. The
mass-loss rate is based on C$_2$ and CN absorption lines and may be
overestimated if the mass loss occurs preferentially along the line of
sight. The extreme mass-loss rate extends the photosphere and lowers
the effective temperature; Bakker et al. suggest that the actual
temperature of the star is much higher than derived from its spectral
type. From the similarity with HD 101584 (section 5.4) Bakker et
al. argue for a binary companion, and narrow chromospheric emission
lines are interpreted as evidence for an accretion disk. They argue
that such an accretion disk could be the source for the high-velocity
outflow.

The OH maser spectra have total velocity widths of $\sim$ 100 km
s$^{-1}$, similar to that of the C$_2$ and CN absorption lines. At
1612 MHz, the OH spectrum is dominated by the strongest emission
feature near 0 km s$^{-1}$ (Fig.\ 5a). The relative strength of this
feature strongly suggests maser amplification of radio continuum
emission which originates nearer the star. The origin of this central
radio continuum source is not clear, since the central star is not hot
enough to ionize the surrounding medium.  We take the position of the
strongest blue-shifted component as the position of the star. With
this choice, the VLA 1612 MHz maser positions of the faint emission at
velocities above 10 km s$^{-1}$ suggest a shell structure (Fig.\ 5a,
middle panel).

\begin{figure}
\psfig{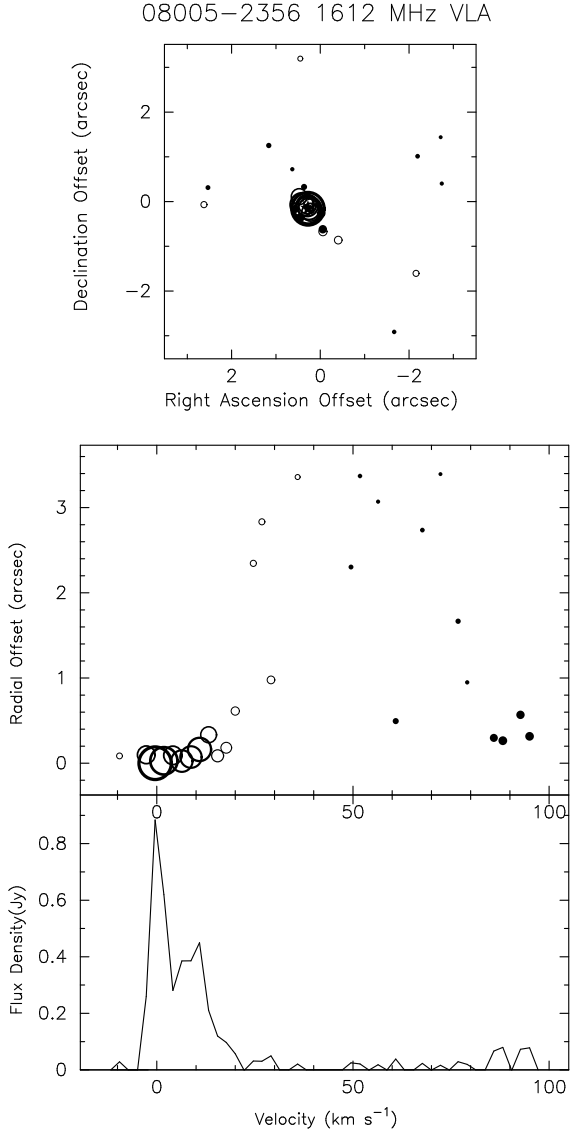}
\caption{(a) IRAS 08005-2356:  $x$-$y$, $R$-$V$ and $I$-$V$
diagrams for the OH 1612 MHz maser emission, from VLA
observations taken in 1990.
}
\end{figure}

\setcounter{figure}{4}
\begin{figure}
\psfig{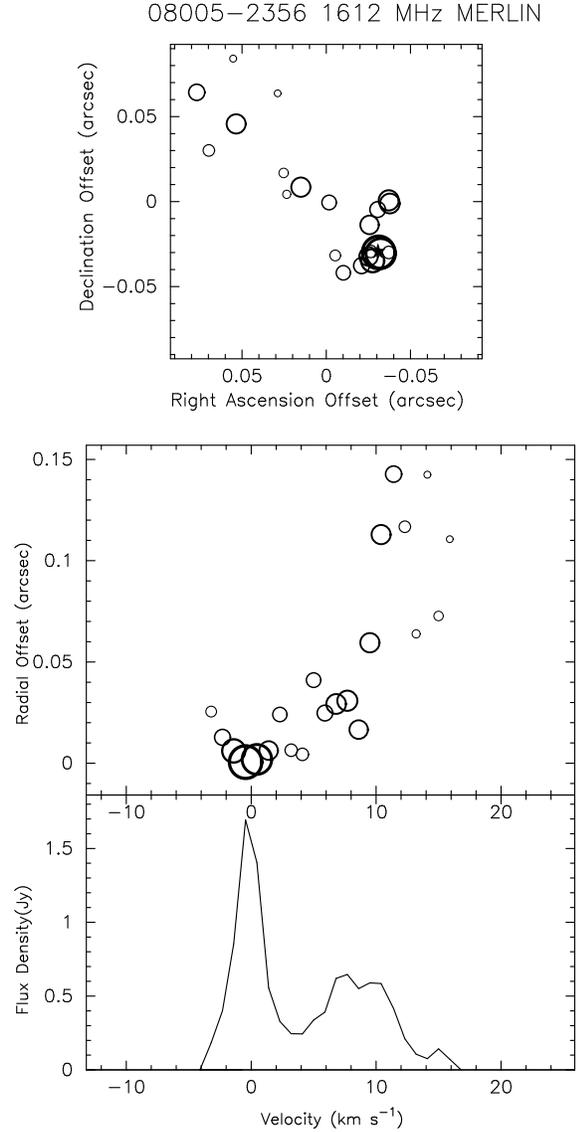}
\caption{(b) IRAS 08005-2356: $x$-$y$, $R$-$V$ and $I$-$V$
diagrams for the OH 1612 MHz maser emission, obtained from MERLIN
observations taken in 1988.
}
\end{figure}

At the extreme blue-shifted velocities the much brighter emission is
evident as a linear $R$-$V$ structure, at almost constant position,
possibly slightly offset in position from the extreme red-shifted
emission.  The linear $R$-$V$ structure is unlikely to correspond to a
true linear outflow.  Linear $R$-$V$ structures at the extreme
velocities are seen in classical OH/IR stars where the emission is
under-resolved (e.g.  Herman 1983; Chapman 1985). The Gaussian fitting
procedures introduce a cutoff velocity, beyond which all components
are centred at the same position. An unresolved structure extending
over an extended velocity range might also occur if there are radial
velocity gradients across the shell, or multiple shells, so that maser
emission is detected over a range of velocities along the
line-of-sight to the star. The higher spatial resolution of the MERLIN
data (Fig.\ 5b) resolves the structure into a shell-like
component. The increase in radius of 0.15 arcsec between 0 and 16 km
s$^{-1}$ is surprisingly small when compared to the maximum offset of
around 3.2 arcsec evident in the VLA data at velocities near 50 km
s$^{-1}$.  Whatever the cause, the small angular scale emission seen
at the extreme blue-shifted velocities is probably physically
different from the linear outflow structures we are investigating in
this paper.

\setcounter{figure}{4}
\begin{figure}
\psfig{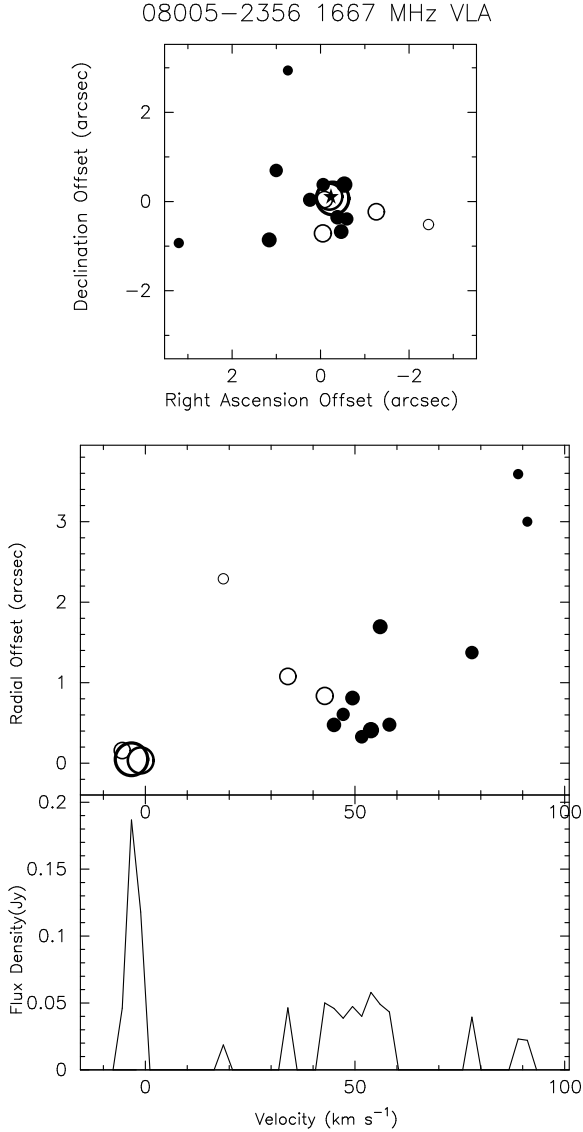}
\caption{(c) IRAS 08005-2356:  Diagnostic $x$-$y$, $R$-$V$ and $I$-$V$ 
diagrams for the OH 1667 MHz maser emission, obtained from VLA
observations taken in 1990.
}
\end{figure}

Fig.\ 5c shows the results obtained with the VLA for the 1667 MHz
masers.  We find that the (fainter) 1667 MHz emission at 0 km s$^{-1}$
is spatially coincident with the extreme blue velocities at 1612
MHz. However the spatial structure of the 1667 MHz masers is not clear
and we do not attempt to interpret this further.

{}From the OH 1612 MHz data, we interpret IRAS 08005$-$2356 as an
expanding shell with an expansion velocity of 45 km s$^{-1}$ and a
shell radius of $3.2^{\prime\prime}\pm 0.2^{\prime\prime}$.  The
expansion velocity of the shell is too high for a normal AGB wind
(e.g. Habing et al. 1994), suggesting acceleration after the original
AGB wind.

\subsubsection{Chemical dichotomy in IRAS 08005$-$2356}

The absorption spectrum of Bakker et al. (1997) indicates a
carbon-rich outflow: it is the only object to show a Phillips
absorption band {\it and} OH emission.  Bakker et al. suggest a recent
change occured from an oxygen-rich to a carbon-rich star.  IRAS
08005$-$2356 does not show the 21-$\mu$m feature which is normally
found in carbon-rich post-AGB stars (Kwok, Hrivnak \&\ Geballe 1995;
Szczerba et al. 1997), supporting a radial gradient in chemical
composition.

The timescales involved suggest  that an old disk acted as an
oxygen-rich reservoir (see the discussion on Roberts 22 which also
shows a chemical dichotomy). The close agreement between the velocities
of the OH masers and the (present) carbon-rich wind of the central
star is interesting, since the chemistry indicates they trace
different gases. However, the carbon wind velocity may have been
underestimated since the absorption lines only measure the line of
sight to the star.

\subsection{Roberts 22 (IRAS 10197$-$5750)} 

Roberts 22 was first discovered as an emission-line object and
classified as a suspected WR star (Roberts 1962). It was later shown
to be a reflection nebula, with the central star completely obscured
but a spectral type of A2I was inferred from the scattered light from
the two lobes (Allen, Hyland \&\ Caswell 1980).  The H$\alpha$ line
arising from the central star has a width of 450 km s$^{-1}$.  Roberts
22 is a strong source of variable OH maser emission at 1612 and 1665
MHz with weaker emission at 1667 MHz.  HST images of this bipolar
nebula (Sahai et al. 1999a) show a dark lane across the centre of the
object, with additional dark regions in the Northern half, close to
where the dark lane is least clear.  Sahai et al.  suggest that the
dark regions are remnants torn off of the torus. The diameter of the
optical nebulosity is $10''\times4''$; the disk is evidently seen
edge-on.

\begin{figure}
\psfig{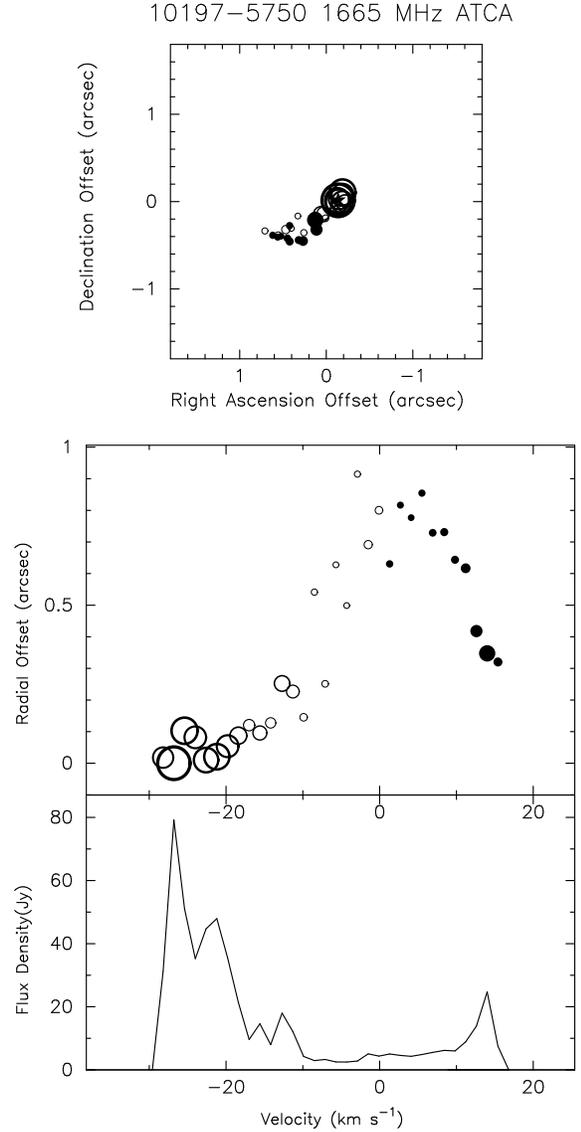}
\caption{(a) Roberts 22: $x$-$y$, $R$-$V$ and $I$-$V$  diagrams for
the OH 1665 MHz maser emission, from ATCA observations in 1992.
observations taken in 1990. The stellar position is assumed to
coincide with strongest blue-shifted 1665 MHz feature at -26.8 km
s$^{-1}$. In Figs 5a--c, the plot origin is chosen to be the
unweighted emission centroid of all the detected maser positions.
}
\end{figure}

\setcounter{figure}{5}
\begin{figure}
\psfig{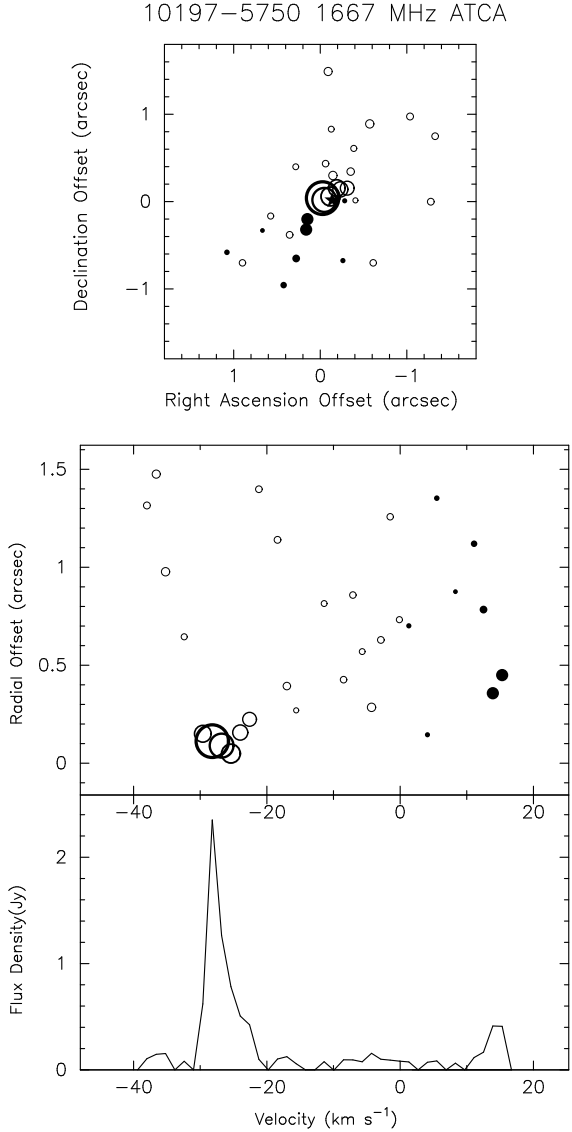}
\caption{(b) Roberts 22:  $x$-$y$, $R$-$V$ and $I$-$V$  
diagrams for the OH 1667 MHz maser
distribution, from ATCA observations in 1992.
}
\end{figure}

Fig.\ 6a shows the diagnostic OH 1665 MHz diagrams obtained from the
ATCA data in 1992. As the star has not been optically identified, the
stellar position is not known to a high precision. We assume that the
stellar position coincides with the 1665 MHz blue-shifted emission
peak at $-$27 km s$^{-1}$. The absolute position of this component is
$\alpha = \rm 10^h 21^m 33.917^s$, $\delta = \rm -58^o 05^\prime
47.730^{\prime\prime}$ (J2000). This is
offset from the centre of the HST nebula, but the phase calibrator
used at the ATCA may have had a slightly uncertain position.

\setcounter{figure}{5}
\begin{figure}
\psfig{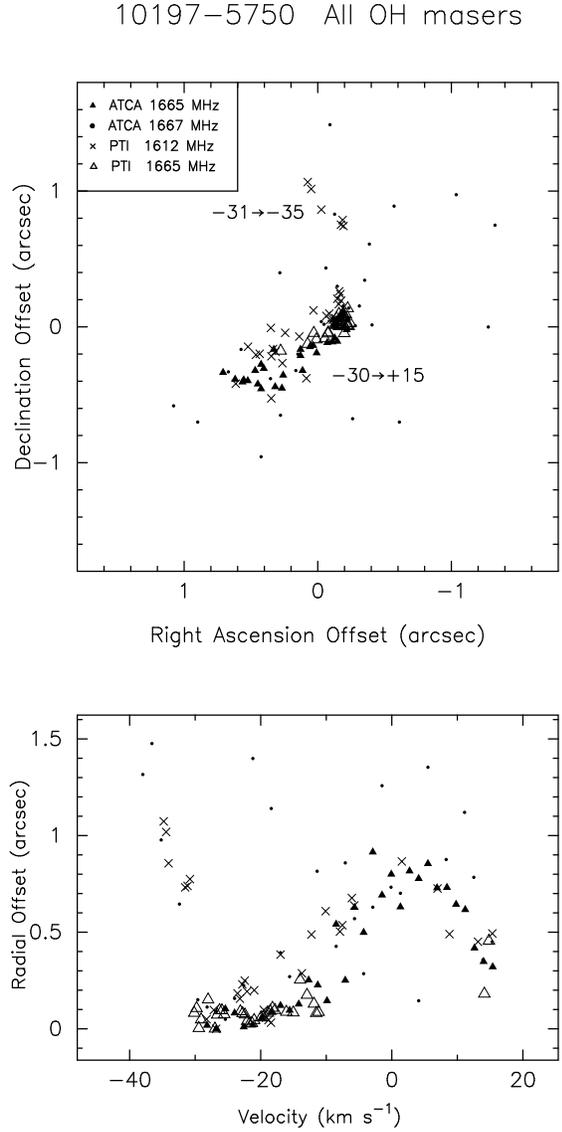}
\caption{(c) Roberts 22:
 Composite $x$-$y$ and
$R$-$V$ diagrams showing the ATCA results in Figs 5a and 5b together
with higher resolution 1612 and 1665 MHz data obtained with the
PTI in 1986/7. The adopted stellar
position, corresponding to the position of the 1665 MHz emission peak
has coordinates of (0.14,0). The velocity ranges of the strong
torus-like structure and the northern feature described in the text
are labelled.}
\end{figure}

At velocities between $-$30 and +15 km s$^{-1}$, the strong 1665 MHz
OH emission from Roberts 22 reveals a torus-like structure of extent
$\sim$1 arcsec which is seen nearly edge on, aligned with the dark
lane seen in the optical images\footnote{For Roberts 22, we also
observed the 1665 MHz emission between -35 and -30 km s$^{-1}$ but are
unable to include that data as it was strongly affected by spectral
ringing from the sharp-edged emission peak at -27 km s$^{-1}$}. From
the maximum observed radial offset at $\sim$ 0.7 arcsec, we take the
stellar velocity to be at 0 km s$^{-1}$. The expansion velocity of the
torus, determined from the blue-shifted emission is approximately 20
km s$^{-1}$. The maser emission from the torus is far brighter on the
blue-shifted side while the extreme velocity red-shifted emission is
not detected.  The gap of emission at the most red-shifted velocities
in Fig.\ 6a can be explained if there is a central ionized region near
the star which is optically thick at 18 cm.

The OH 1665 MHz distribution of Roberts 22 is similar to the OH 1612
MHz distribution of IRAS 08005$-$2356 (Section 5.2). Both sources also
show a horizontal feature in the $R$-$V$ diagram which extends over
$\sim$ 10 km s$^{-1}$ at the outer blue-shifted velocities. As
discussed for IRAS 08005-2356, this feature is unlikely to correspond
to a true linear outflow.  The weaker 1667 MHz emission of Roberts 22
(Fig.\ 6b) covers a larger area of $\sim$ 2.5 arcsec. As for other
sources, the maser geometry is least clear at 1667 MHz. Within the
position errors the peak emission at 1665 and 1667 MHz is coincident.

Fig.\ 6c shows composite $x$-$y$ and $R$-$V$ diagrams where the ATCA
1665 and 1667 MHz results are plotted together with the higher angular
resolution 1665 and 1612 MHz results from the PTI observations. We
have aligned the PTI positions with the ATCA positions by assuming
that the positions of the 1612 and 1665 MHz emission features at $-27$
km s$^{-1}$ are coincident.

The PTI maser positions and velocities confirm the torus-like
structure evident from the ATCA 1665 MHz data, for the velocity range
$-30$ to 15 km s$^{-1}$. At 1612 MHz, the PTI observations also
detected the weaker emission feature at the extreme blue-shifted
velocities between $-31$ and $-35$ km s$^{-1}$. This feature is
evident as the northern group in Fig.\ 6c. Across this group there is
a systematic velocity gradient from $-31$ km s $^{-1}$ at the southern
edge to $-35$ km s$^{-1}$ at the northern edge. In the $R$-$V$ diagram
the northern feature is seen as a linear feature which is aligned with
part of the 1667 MHz emission. The 1665 MHz PTI data confirms the large
range of velocities (15 km s$^{-1}$) seen in the blue-shifted emission
towards the central star. These velocities may trace a radial velocity
gradient or may be due to a turbulent velocity field.

Overall we interpret the maser emission from Roberts 22 as a nearly
edge-on disk which coincides with the dark lane between the two
optical reflection lobes, as expected if the torus is oxygen-rich. The
detection of the linear feature to the north and the more scattered
1667 MHz emission indicates that some collimated outflows may also be
present.  

\subsubsection{Chemical dichotomy in Roberts 22}

\setcounter{figure}{5}
\begin{figure}
\psfig{figure=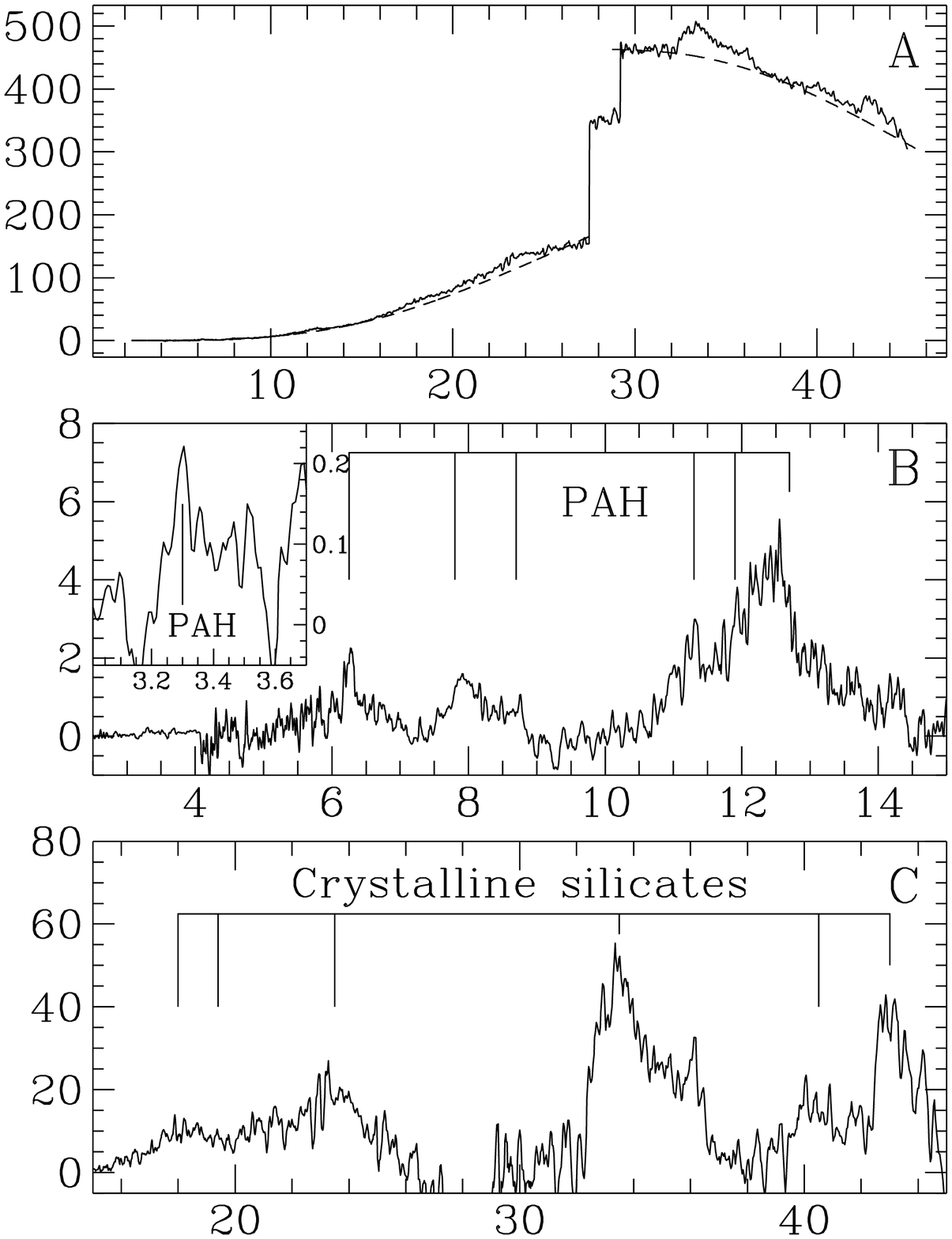,width=90mm}
\caption{(d) The SWS spectrum of Roberts 22. In panel A the
complete SWS spectrum is shown, together with the spline fit
continuum. In panel B the continuum-subtracted spectrum up to 15$\mu$m
is shown with the PAH-features indicated.  Panel C shows the
continuum-subtracted spectrum longwards of 15 $\mu$m; indicated are
the positions of crystalline silicate features.  The spectra were
reduced using the SWS off-line processing software, version 7.0.
Fringes in the 16.--29.2$\mu$m part of the spectrum were removed using
the InterActive (IA) data reduction package routine {\em
fringes}. Flux and wavelength calibration procedures are described in
Schaeidt et al. (1996) and Valentijn et al. (1996).
}
\end{figure}

Molster et al.\ (1996) have suggested that Roberts 22 exhibits a mixed
oxygen/carbon chemistry. The ISO (Infrared Space Observatory, Kessler
et al. 1996) spectrum of Roberts 22 is shown in Fig.\ 6d.  The
observations were taken on July 27, 1996, with the Short Wavelength
Spectrometer (SWS, de Graauw et al. 1996).  The total integration time
was 3454 seconds. Unfortunately, ISO was pointed about
$11^{\prime\prime}$ of the centre, and the flux calibration is badly
affected by wavelength-dependent aperture losses. It also caused flux
jumps at 27.5 and 29.2 $\mu$m because of the larger apertures used at
these wavelengths. The flux calibration is therefore not reliable.

Bad data points were identified by comparing different detectors at
the same wavelength, and were manually removed. The detectors were
combined into a single spectrum for each of the 12 subbands.  The
different sub-band spectra were scaled to match each other for band~1A
to 3D (2.4 -- 27.5 $\mu$m).  Band~3E (27.5 -- 29.2 $\mu$m) and band~4
(29.2 -- 45 $\mu$m) are not scaled to the other detectors: the jumps
are caused by the pointing error.

To enhance the features present on top of the continuum, we have
subtracted a spline (see Fig.\ 6d) fitted through selected points
where no feature was believed to be present, separately for band 1A to
band 3D and for band 4. For band 3E no continuum was drawn because of
the short wavelength coverage in this band.  Very broad features are
treated as continuum by this method, and the strength of broad
features may be reduced.  The spline fit continuum has no physical
meaning and is only used to enhance the features.

The dual chemical character of the dust is clearly visible in the
continuum-subtracted spectrum. Carbon-rich dust is indicated by the
PAH features at 3.3, 6.25, 7.8, 8.7, 11.3, 11.9 and 12.7 $\mu$m.
However, at wavelengths longwards of 15 $\mu$m the crystalline
silicates features at 18, 19.4, 23.5, 33.5, 40.5 and 43 $\mu$m
dominate the spectrum. There is also a hint of the broad 18 $\mu$m
amorphous silicates feature. The crystalline silicate feature at
43$\mu$m is probably blended with crystalline water-ice, since there
is a hint in the LWS (45 -- 200 $\mu$) spectrum (not shown) of a 60
$\mu$m bump, which is usually attributed to crystalline water-ice (see
e.g. Barlow 1998).  All these are a clear sign of the presence of
oxygen-rich dust species.  The strength of the crystalline silicate
features compared to the continuum is within the range of ordinary
outflow sources (Molster et al. 1999).

This chemical dichotomy, with both carbon-rich and oxygen-rich dust
being simultaneously present, is shown by only a few post-AGB stars:
the binary (post-)AGB nebula `the Red Rectangle' (HD 44179: Waters et
al. 1998a), a number of planetary nebulae with WR central stars
(so-called [WC] stars) (Waters et al. 1998b), and the extreme bipolar
PN NGC 6302. Waters et al. assume that the silicate features arise
from an old disk, formed at a time when the central star was
oxygen-rich (see also Jura \&\ Kahane 1999), and the present outflow
is carbon rich.  Of the AGB carbon stars which show silicate emission
(Willems \&\ de Jong 1986, Little Marenin 1986), only two show
evidence for both silicate and carbon (SiC) circumstellar features:
IRAS 04496$-$6958 in the LMC (Trams et al. 1999) and CS1003 in the
Galaxy (Little Marenin 1986).  The other silicate carbon stars show
only oxygen-rich dust. The preferred explanation for the silicate
carbon star also involves a long-lived disk (Lloyd Evans 1990).

\subsection{HD 101584 (IRAS 11385$-$5517)} 

HD 101584 is an optically identified star of spectral type B9II
(Bakker et al.\ 1996a) and belongs to a group of high-latitude,
bright, A--G post-AGB stars (Bidelman 1951; Parthasarathy \& Pottasch
1986).  From optical photometry, Bakker et al.\ (1996b) obtained a
periodicity of 218 days and inferred that the star is in a close
binary system with an unseen  companion.  The star has a
strong stellar wind with a maximum outflow velocity close to the star,
determined from Balmer line P Cygni profiles, of $\sim$ 100 km
s$^{-1}$ (Trams et al.\ 1990; Bakker et al.\ 1996b).
The distance is in the range 0.6--1.1 kpc (Bakker et al. 1996a).
The Hipparcos parallax is $1.32 \pm 0.77$ milli-arcsec.

Molecular $^{12}$CO and $^{13}$CO emission from HD 101584 occurs over
an extreme velocity range of almost 300 km s$^{-1}$ (Trams et
al. 1990; Loup et al. 1990; Van der Veen, Trams \&\ Waters 1993;
Olofsson \& Nyman 1999). The $^{12}$CO spectral profiles show weakly
double-peaked emission which extends over $\sim$ 100 km s$^{-1}$,
together with highly extended emission wings which cover $\sim$ 300 km
s$^{-1}$ and two relatively narrow features at velocities of $-92$ and
173 km s$^{-1}$. The stellar velocity is well defined from the centre
of the $^{12}$CO profiles to be $41\pm2$ km s$^{-1}$ (Olofsson \&
Nyman 1999) The $^{13}$CO profiles are similar but show an additional
narrow feature which is centred precisely at the stellar velocity.

Olofsson \& Nyman find that the $^{12}$CO(2$\rightarrow$1) emission is
extended along a position angle of 90$^0$. The high velocity CO
emission is bipolar with higher expansion velocities detected at
larger distances from the star. The extreme blue and red-shifted CO
emission is offset by $\sim$ 4.5 arcsec to the west and east of the
stellar position respectively with an average CO velocity gradient
between the two sides of approximately 30 km s$^{-1}$
arcsec$^{-1}$. The spatially unresolved low and intermediate velocity
CO emission within $\sim$ 60--70 km s$^{-1}$ of the stellar velocity
is located much closer to the star within a region of radius $\sim$ 1
arcsec, indicating a higher velocity gradient or turbulence in this
region.

\begin{figure}
\psfig{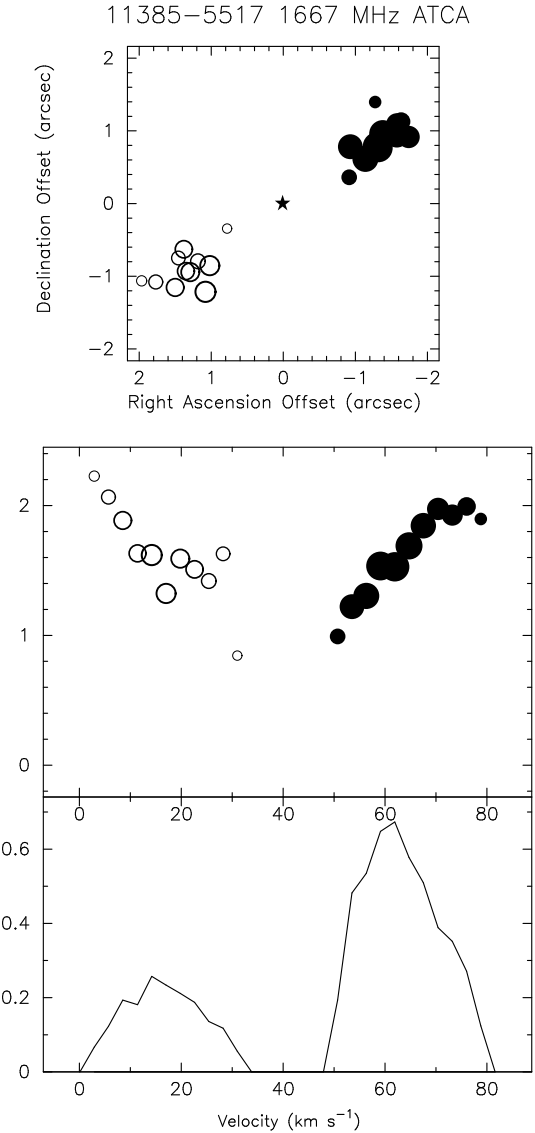}
\caption{(a) HD 101584:  $x$-$y$, $R$-$V$ and $I$-$V$ diagrams for the
OH 1667 MHz maser emission, obtained from ATCA observations taken in
1991. 
}
\end{figure}

The OH 1667 MHz maser distribution for HD 101584 has previously been
discussed by te Lintel Hekkert, Chapman \& Zijlstra (1992) but is
included here for a comparison with the recent CO results. The OH 1667
MHz emission (Fig.\ 7a) shows two emission features over a
considerably smaller total velocity range of 84 km s$^{-1}$. The
central OH velocity agrees well with the systemic CO velocity.  The
masers are located within two groups at a position angle of
$-$60$^{\rm o}$ which extend between one and two arcseconds on either
side of the star. The systematic velocity gradient is approximately 25
km s$^{-1}$ arcsec $^{-1}$.

\setcounter{figure}{6}
\begin{figure}
\psfig{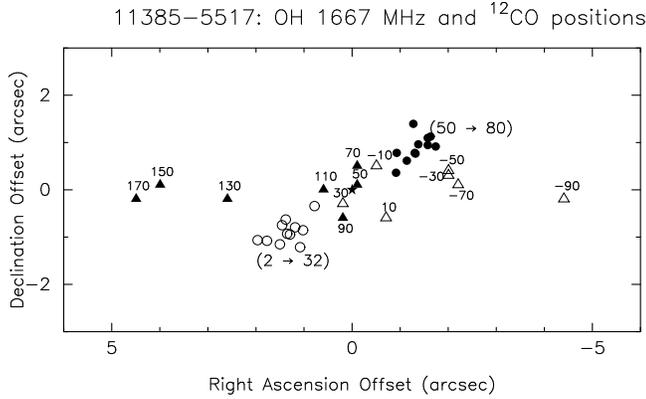}
\caption{(b) HD 101584:ATCA OH 1667 MHz maser positions plotted together
 with the
$^{12}$CO$J=2\rightarrow1$ positions from SEST observations in 1997
(Olofsson \& Nyman 1999). The velocity at each CO position is
labelled. The velocity ranges for the two groups of OH masers are
given in brackets. The stellar position is assumed to be at the
centroid of both the OH and CO distributions as indicated by the
filled star at the origin.
}
\end{figure}

Fig.\ 7b shows the ATCA maser positions plotted together with the CO
positions from Olofsson \& Nyman (1999). Olofsson \& Nyman have
presented a geometrical model in which the OH maser emission together
with the low and intermediate velocity CO emission are located within
an equatorial disk while the high velocity CO emission originates in a
bipolar outflow aligned in the perpendicular polar
directions. However, from the model presented in this paper, the
linear velocity gradient of the OH masers is consistent with a bipolar
outflow.  No evidence for a disk- or shell-like OH structure is seen.

The narrow $^{13}$CO emission feature with an expansion velocity of 7
kms$^{-1}$, centred on the stellar velocity, shows evidence for a
molecular disk near the star. The lack of any OH emission from this
disk can be understood if the density in the disk is sufficiently high
for collisional de-excitation of the OH molecules. The absence of the
disk feature in the $^{12}$CO line is consistent with a high optical
depth and high density in material surrounding the disk. Both the
intermediate velocity CO and OH may come from this surrounding gas.

Although the averaged velocity gradient of the OH masers is similar to
that of the high velocity CO emission, the bipolar OH distribution
(observed in 1991) cannot be aligned with the high velocity bipolar CO
distribution (observed in 1997).  Fig. 7b\ shows that the east-west
axis of the CO emission is rotated by an angle of $\sim$ 150$^{\rm o}$
(anticlockwise) with respect to the northwest-southeast OH axis.  The
OH and CO therefore trace {\it different} components. The CO is likely
to trace the most massive component. The OH could come from a smaller
region of higher excitation, such as e separate jet-like structure
which developed within the expanding gas (e.g., Frosty Leo: Sahai et
al. 2000).  Another possibility, also raised by Olofsson \& Nyman, is
that some precessional motion is present.

Although Bakker et al.  suggest that the disk is seen edge on, a
moderate inclination as suggested by Olofsson \&\ Nyman seems
consistent with the available data.  HD 101584 is associated with an
optically bright post-AGB star which indicates a low circumstellar
extinction. The CO velocity gradient also argues against either a
pole-on or edge-on viewing angle.  The observed velocity variations
arising from the binary orbit indicates that the line of sight cannot
be close to pole-on, unless the circumstellar disk and the binary
orbit are in different planes.

\subsection{IRAS 15405$-$4945} 

Single-dish OH spectra of IRAS 15405$-$4945 (te Lintel Hekkert \&\
Chapman 1996; te Lintel Hekkert et al. 1988) show a broad velocity
range of 80 km s$^{-1}$ at 1612 MHz, 105 km s$^{-1}$ at 1665 MHz and
155 km s$^{-1}$ at 1667 MHz.  No optical counterpart is known. Due to
its location in the Galactic plane and cold IRAS colours, the object
has been interpreted as a possible ultra-compact HII region. A search
for methanol maser emission was unsuccessful and so did not confirm
such a classification (Walsh et al. 1997). From the centre of the two
main-line OH profiles, we take the stellar velocity to be 60 km
s$^{-1}$. The centre of the 1612 MHz profile is offset, with a median
value of 74 km s$^{-1}$.

\begin{figure}
\psfig{figure=fig8a.ps,height=150mm}
\caption{(a) IRAS 15405$-$4945:  $x$-$y$, $R$-$V$ and $I$-$V$ diagrams
for the OH 1665 MHz maser emission, from ATCA observations taken in
1991.
}
\end{figure}
\setcounter{figure}{7}
\begin{figure}
\psfig{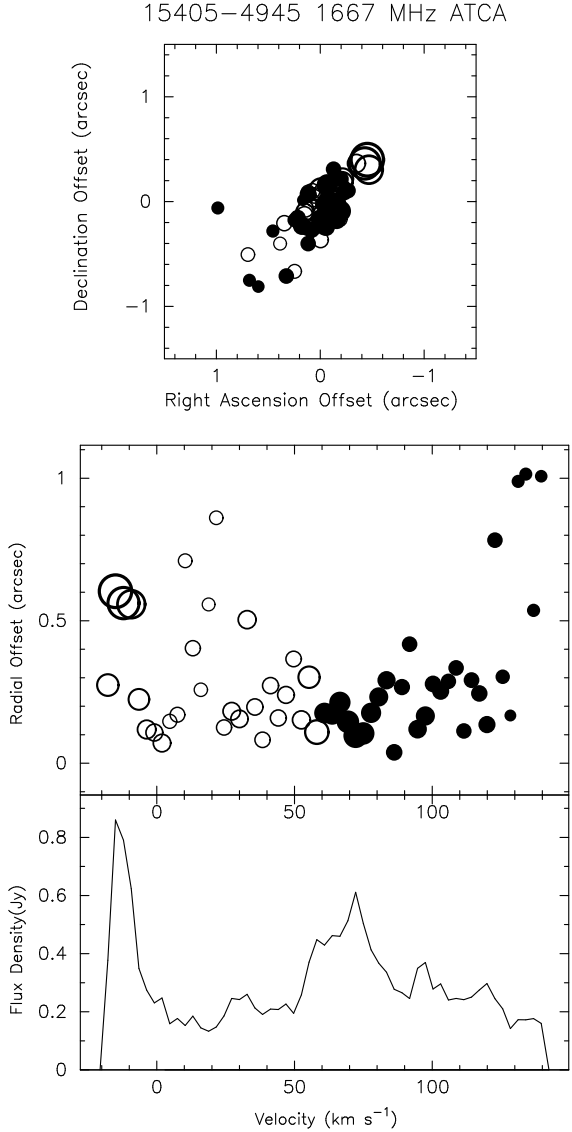}
\caption{(b) IRAS 15405$-$4945:   $x$-$y$, $R$-$V$ and $I$-$V$ diagrams for 
the OH 1667 MHz maser emission, from ATCA observations taken in 1991.
}
\end{figure}

The OH 1665 and 1667 MHz $x$-$y$ plots (Figs 8a, b) show that the
masers lie in an elongated elliptical-shaped region, with a position
angle of $-$45 degrees and a total diameter of approximately 2
arcsec. At 1665 MHz there is a clear separation of the blue- and
red-shifted masers. The stellar position is assumed to coincide with
the unweighted emission centroid of the OH 1665 and 1667 MHz maser
positions. te Lintel Hekkert et al.\ (1988) attribute the extreme
blue-shifted feature to amplification of the stellar continuum; the
present data does not support this. 

The separation of the blue and red-shifted masers at 1665 MHz argues
for an axi-symmetric, possibly bipolar outflow which is however very
poorly resolved (Fig.\ 8b).  At 1667 MHz the largest offsets occur at
the extreme velocities, more consistent with a linear outflow than
with an expanding shell.  The extreme velocity ranges of the OH
spectra are not easily explained with an expanding shell structure. We
therefore interpret the OH maser spectra and positions as evidence for
a wind--wind interaction with `jet-like' rather than `shell-like'
maser distributions. Higher spatial resolution observations are needed
to confirm this hypothesis.

The OH profiles show significant variability.  The OH 1612 MHz maser
profile has nearly doubled in strength overall since its discovery in
1987, whilst a large number of individual maser spikes have appeared
and others have disappeared (see Figs 1 and 2, te Lintel Hekkert et
al.\ 1988). The OH 1667 MHz feature between --20 km s$^{-1}$ and --10
km s$^{-1}$ has decreased significantly in strength and other, broad
features have increased.

\subsection{IRAS 16342$-$3814} 

This outflow source was first discovered by Likkel \&\ Morris (1988)
and te Lintel Hekkert et al.  (1988), and shows both OH and H$_2$O
outflows. The highly variable H$_2$O masers occur in pairs at opposite
velocity, at a number of discrete velocities (Likkel, Morris \&\
Maddalena 1992).  HST images show two asymmetric bipolar reflection
lobes on either side of a dark lane which is identified as an
edge-on equatorial disk or torus (Sahai et al. 1999b). The
forbidden stellar velocity argues that IRAS 16342$-$3814
belongs to a lower-mass, old population.

\begin{figure}
\psfig{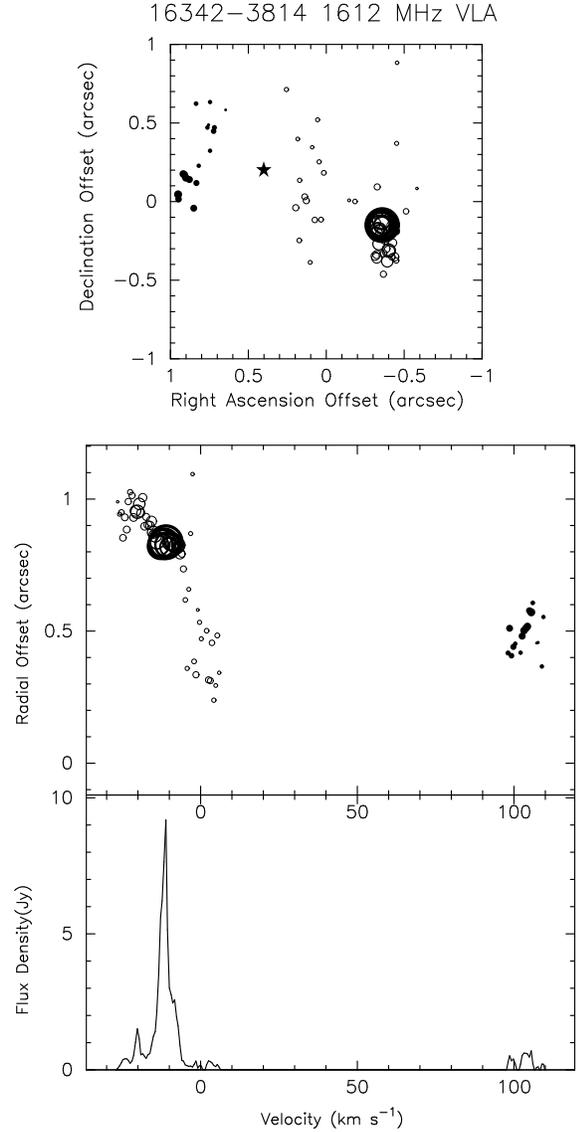}
\caption{(a) IRAS 16342$-$3814:   $x$-$y$, $R$-$V$ and $I$-$V$ diagrams
for the OH 1612 MHz maser emission, from VLA observations taken in
1990.
}
\end{figure}

Fig.\ 9a shows the maser results obtained from the VLA 1612 MHz
observations.  The red- and blue-shifted masers are located in two
well-separated groups with an east-west gap between the two groups of
$\sim$ 0.35 arcsec which coincides with the location of the central
torus. From the $R$-$V$ diagram, the blue-shifted masers show a linear
structure which extends over $\sim$ 0.8 arcsec, similar to HD 101584.

\setcounter{figure}{8}
\begin{figure}
\psfig{figure=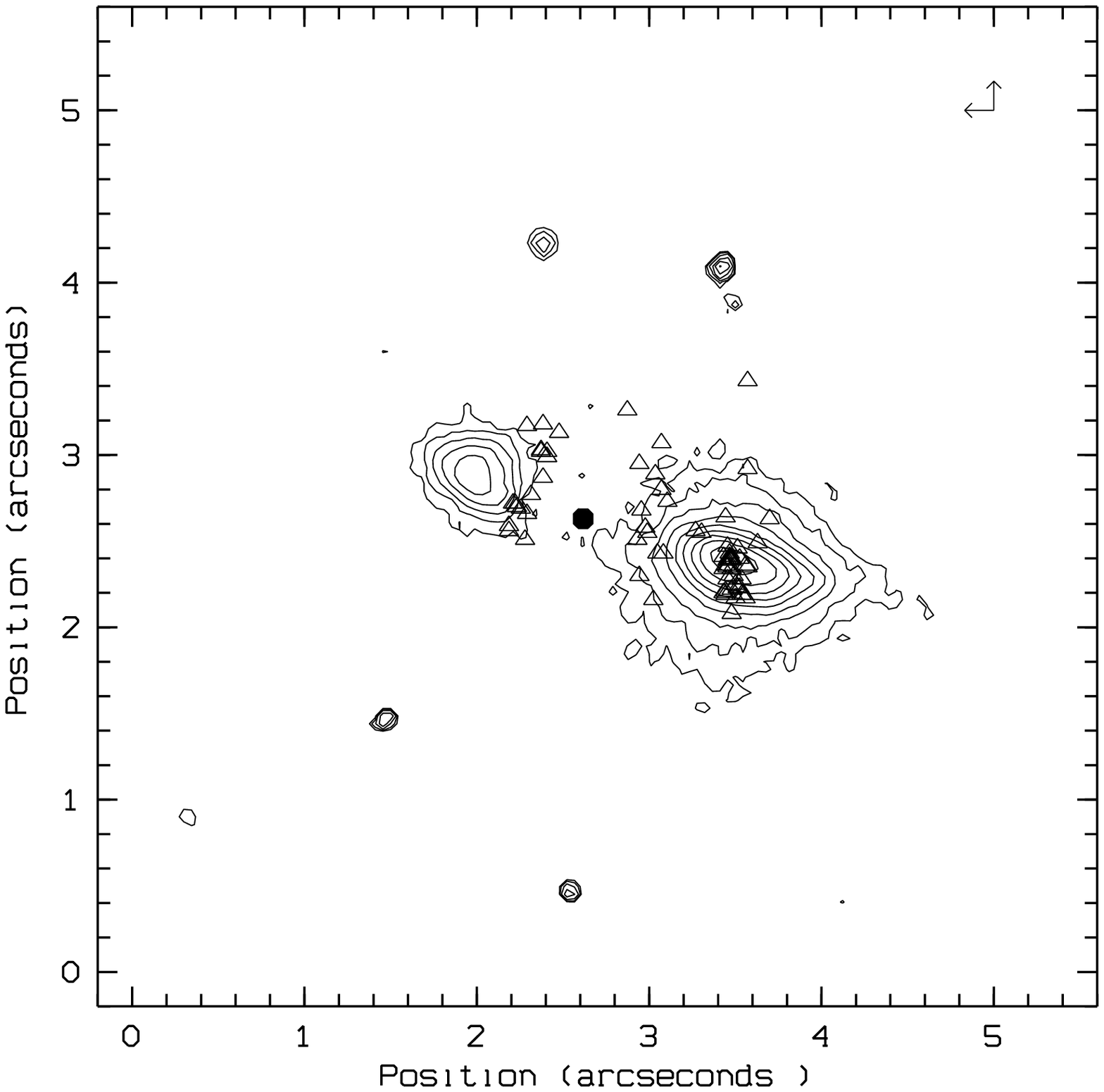,width=85mm}
\caption{(b) IRAS 16342$-$3814: An HST image through the V-filter (F555) 
taken with the
Planetary Camera in 1997, calibrated using the HST pipeline
software. North is up and east is to the left. The pixel scale is
0.045 arcsec.  The image has been de-rotated and corrected for field
distortions. The contour levels are in steps which increase by a
factor of 2.  The triangles indicate the positions of the OH
components. The filled star shows the estimated position of the
central star, at $\rm \alpha=16^h34^m17.06^s$, $\rm
\delta=-34^o18^\prime17.9^{\prime\prime}$ (B1950).  The error bars in
the corner indicate the uncertainty in alignment of radio and optical
coordinates.
}
\end{figure}

We recalibrated an archive HST V-band image of IRAS 16342$-$3814
(principal investigator Bobrowski; used in the V--I colour composite
image of Sahai et al. 1999b). An astrometric accuracy of $\sim$ 0.2
arcsec was achieved using the Digital Sky Survey, and the frame was
rotated and corrected for astrometric distortions.  Fig.\ 9b shows the
OH 1612 MHz maser positions overlaid on the HST image. The red-shifted
masers are detected from the inner edges of the eastern lobe while the
stronger blue-shifted masers are located along the inner edge and
towards the centre of the brighter western lobe.  The OH masers avoid
the dark torus. The strongest OH emission feature, at $-11$ km
s$^{-1}$, coincides with the brightest point of the reflection
nebula. This is almost certainly due to amplification of radio
continuum emission from the lobe. Radio continuum emission may arise
from shock-ionization due to the presence of a narrow jet within the
lobe, as in  M1-92 where the radio continuum emission is
also centred on the brightest reflection lobe rather than at the star
(Bujarrabal et al.\ 1998).

The OH spectra of IRAS 16342$-$3814 have similar variability
characteristics to IRAS 15405$-$4945 (see Figs 3 and 4, te Lintel
Hekkert et al.\ 1988).  Whilst the OH 1612 MHz emission was at a
similar strength in 1988 and 1990, the OH mainline emission has
increased in intensity by more than 50 per cent with striking changes
in the spectral profiles. One difference between this source and IRAS
15405$-$4945 is that the OH maser emission is strongly circularly
polarized in all lines, which is rare in OH/IR stars.

\subsubsection{The disk parameters of IRAS 16342$-$3814}

\setcounter{figure}{8}
\begin{figure}
\psfig{figure=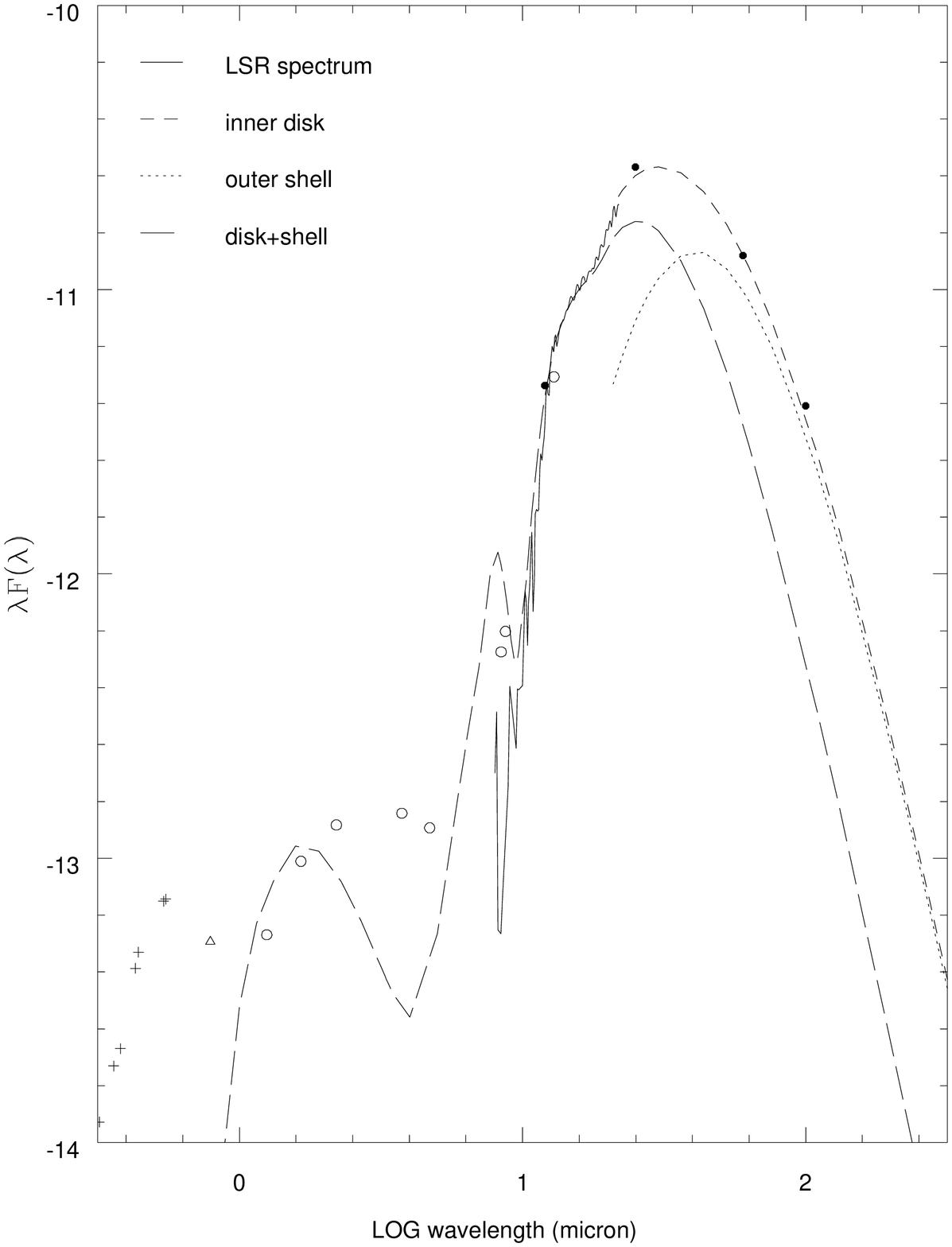,width=85mm}
\caption{(c) The infrared spectrum of IRAS 16342$-$3814, showing
the IRAS LRS spectrum, the IRAS photometry (filled circles), and
optical and infrared photometry (other symbols) of van der Veen et
al. (1988). The optical photometry appears to be wrong, possible
related to a nearby field star.  The fit consisting of star+disk+shell
is indicated, with the shell also shown separately.
}
\end{figure}

The density and size of the edge-on disk of IRAS 16342$-$3814 can be
determined using infrared modeling.  We have recalibrated the LRS
spectrum using the IRAS Software Telescope (Assendorp et al. 1995).
The spectrum plus available photometry (van der Veen et al. 1989) is
shown in Fig.  9c. The 18-$\mu$m feature is in absorption, and a lack
of hot dust is evident shortward of the 9.8-$\mu$m feature.

We fitted the spectrum using a radiative-transfer code (see
Siebenmorgen, Zijlstra \& Kr\"ugel 1994) which includes a range of
dust sizes and incorporates scattering effects. The code assumes
spherical symmetry.  We applied a correction factor to allow for the
fact that the torus only covers a finite solid angle as seen from the
star. This approach is possible because the disk is almost seen
edge-on.  The ratio of optical depth of the 18-$\mu$m to 9.8-$\mu$m
features is very sensitive to the temperature of the innermost dust.
Thus the inner radius of the dust shell is quite well determined. The
best fit is shown in Fig. 9c, using model parameters summarised in
Table 4.

The best fit indicates a  distance  considerable less than
assumed by Sahai et al. (1999b). We used a distance of 700 pc.
At this distance the star is somewhat sub-luminous compared to most
Post-AGB stars, but at larger distances the mass in the shell becomes 
too large for AGB mass loss. The present distance would reduce the
extremely high mass-loss rate implied by Sahai et al. 

The torus parameters required to fit the LRS spectrum indicate an
inner radius of approximately $1^{\prime\prime}$ and an outer radius
of $1.5^{\prime\prime}$. The values agree very well with the HST image
in Fig.\ 9b.  The dust temperature runs from 200K at the inner edge to
80K at the outer. The density of the disk is about $10^8\,\rm
cm^{-3}$: this is far above the density where OH masers are
collisionally de-excited and is in agreement with the fact that no OH
masers are seen towards the dark torus. The torus in the model covers
about 10\%\ of the sky as seen from the central star, and has a mass
of about 0.1 M$_\odot$.

The HST image of IRAS 16342$-$3814 reveals four remarkable point-like
sources (Fig.\ 9b) distributed almost symmetrically around the
obscured torus.  Their positions correspond well with the outer edge
of the disk model. This suggests that they may be caused by scattered
emission from the edge of the disk.

\begin{table}
\begin{minipage}{160mm}
\caption{Infrared fit to IRAS 16342$-$3814}
\begin{tabular}{lcrcrrccccc}
\hline
star & \\
distance & 700 pc \\
$L$ & 2500 L$_\odot$ \\
$T$ & 15000 K \\
\hline
nebula          & disk                    & outer shell \\
inner radius    & $9.5\,10^{15}$ cm       & $5\,10^{16}$ cm\\
outer radius    & $1.5\,10^{16}$ cm       & $4\,10^{17}$ cm \\
inner density   & $1.4\, 10^8$ cm$^{-3}$  & $1\, 10^5$ cm$^{-3}$ \\
density profile & $ r^{-2} $              & $ r^{-1} $\\
covering angle  & 10 \%                   & 90 \% \\
Mass            & 0.10 M$_\odot$          & 4: M$_\odot$ \\
\hline
 \end{tabular}
\end{minipage}
\end{table}

The long wavelength IRAS fluxes cannot be fitted with a disk.  Instead
we had to assume an outer shell. The fit parameters from this shell
are however not well determined. To reduce the mass in this shell, a
shallow density gradient of $r^{-1}$ was chosen (based loosely on
eq. (12)). This dust may be located in the intershell region between
the two winds, in which case the values in Table 4 are not
relevant. The 60-$\mu$m flux may also in part be due to the
crystalline ice band.

The optical photometry of van der Veen et al. cannot be fitted, and
are also inconsistent with the HST image (Sahai et al. 1999b): it is
believed that these data represent a nearby field star. The infrared
data also is not well fitted: the model predicts that the extinction
towards the star ($A_V=27$ mag) is too low to hide the star beyond a
few micron. The fact that no bright star is seen in the infrared
indicates both a low luminosity and a high temperature.  However, the
stellar temperature cannot be much higher than 15 kK since no
ionization is seen.

\subsection{IRAS 17253$-$2831} 

This OH source was listed by Zijlstra et al. (1989) as a possible
coincidence with the planetary nebula Th3-19 (PN358.4+03.3; IRAS
17253$-$2824). However, the present OH data shows that the Parkes
position was offset by $7^\prime$.  The far more accurate VLA position
is within several arcseconds of IRAS 17253$-$2831, which has a very
red 12/25 $\mu$m colour with strongest infrared emission at 25$\mu$m,
consistent with an OH/IR star with deep silicate absorption. A SIMBAD
search did not reveal any other published data of this source.

The OH profile shows peculiar emission wings outside of the two
strongest OH peaks.  OH 1612 MHz emission was detected at velocities
between $-$78.2 and $-$46.4 km s$^{-1}$ with the strongest emission
from the two emission peaks at $-$53.2 and $-$71.4 km s$^{-1}$. From
the mid-velocity of the two peaks we take the stellar velocity to be
at $-$62 km s$^{-1}$.

\begin{figure}
\psfig{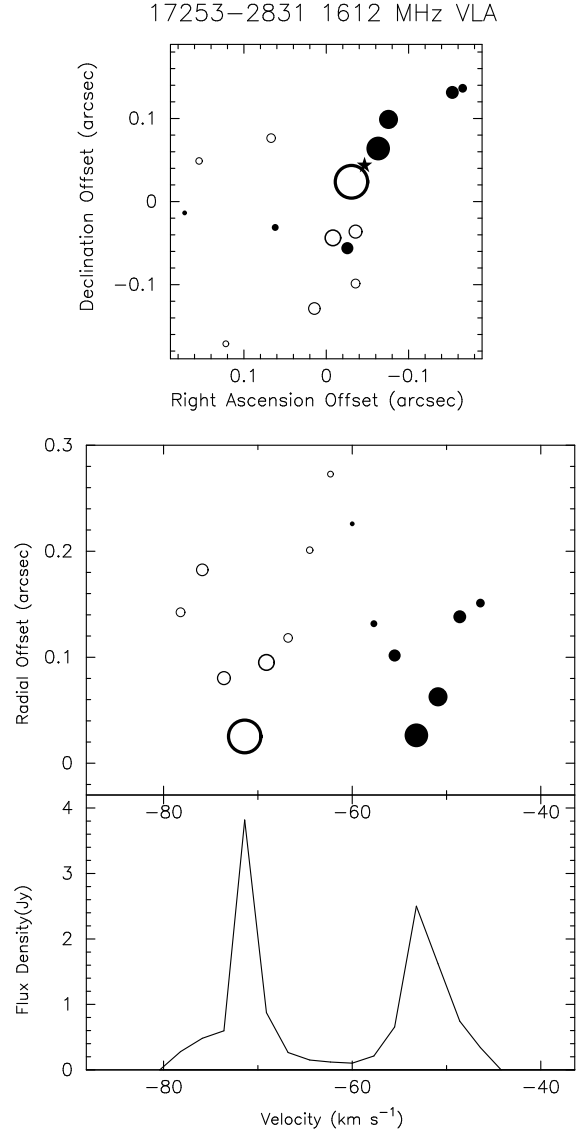}
\caption{IRAS 17253$-$2831:  $x$-$y$, $R$-$V$ and $I$-$V$ diagrams
for the OH 1612 MHz maser emission, from VLA observations taken in 1988.
}
\end{figure}

The OH 1612 MHz maser emission (Fig. 10) was detected within a
compact, poorly resolved region of $0.35$ arcsec diameter. In Fig.\ 10
we have adopted a position mid-way between the positions of the two
emission peaks as the best estimate of the stellar position. For this
stellar position, the $R$-$V$ diagram shows an elliptical distribution
of points at velocities between the two emission peaks as expected for
a circumstellar shell with an expansion velocity of 9 km s$^{-1}$ and
an angular radius of approximately 0.25 arcsec.  In contrast, at
velocities outside of the emission peaks the maser positions show an
increase in the position offsets with velocity, indicating a likely
bipolar outflow.

{}From the VLA data we interpret IRAS 17253$-$2831 as an AGB wind with
low expansion velocity, where the emission profile is extended beyond
the expansion velocity of the shell due to a bipolar outflow which
results from a wind--wind interaction. Higher resolution observations
are needed to confirm this interpretation.

\subsection{IRAS 17423$-$1755 (He3-1475)} 

This object shows  one of the most spectacular bipolar post-AGB
outflows known (Bobrowski et al.\ 1995; Riera et al.\ 1994; Borkowski,
Blondin \&\ Harrington 1997). The optical images show a curved line of
shock-ionized knots on both sides of the central object, with
velocities in excess of 500 km s$^{-1}$. The star is classified as a B[e]
star and is only seen in reflection against the back of the torus
(Bobrowski et al.\ 1995). The emission spectrum shows many iron lines
arising from very dense gas close to the star (Riera et al.\ 1994, Bautista
\&\ Pradhan 1998).

\begin{figure}
\psfig{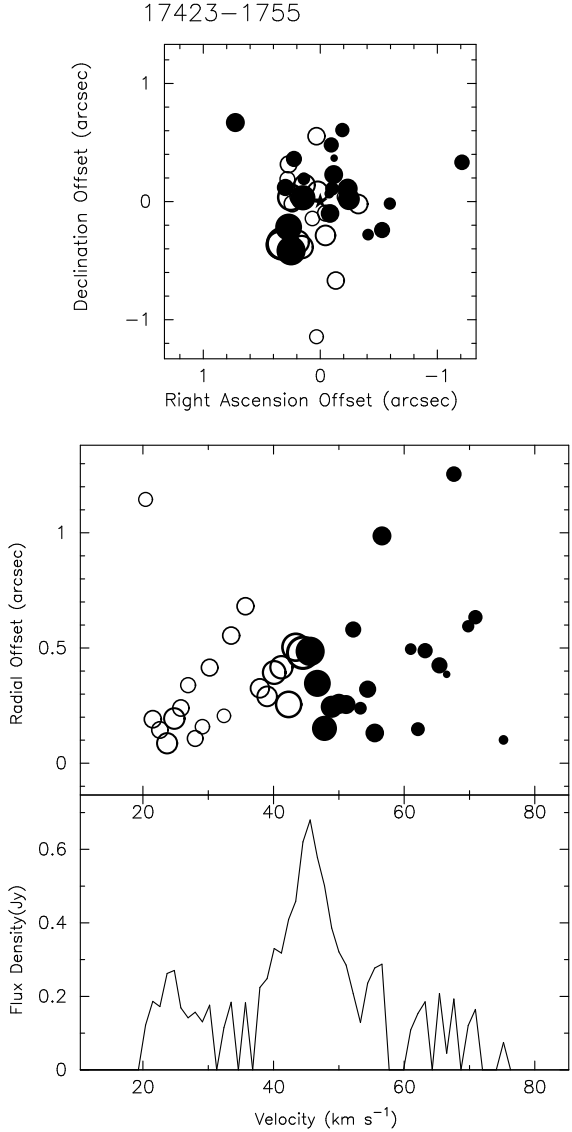}
\caption{IRAS 17423$-$1755:  $x$-$y$, $R$-$V$ and $I$-$V$ diagrams
for the OH 1667 MHz maser emission, from VLA observations taken in 1988.
}
\end{figure}

The OH 1667 MHz spectrum is faint, and shows a fairly flat profile but
with enhanced emission near the stellar velocity at $\sim$ 50 km
s$^{-1}$.  Bobrowski et al.\ attribute this feature to amplification
of radio continuum emission from the shock-ionized jet.  The diagrams
of Fig.\ 11 show a poorly resolved shell-like structure which is
better defined on the blue-shifted side. On the red-shifted side the
emission is more scattered in position but no clear linear outflow is
visible. We interpret the OH as being associated with the expanding
shell or torus.

\subsection{IRAS 18491$-$0207} 

The classification of IRAS 18491$-$0207 is uncertain. The IRAS colours
are consistent with an HII region (Fig.\ 1), however Walsh et al.\
(1998) did not detect methanol maser emission and the NVSS survey did
not reveal radio continuum emission (Condon et al.\ 1988). The IRAS
variability index of 31\%\ is more consistent with a classification as
a post-AGB star.  Its nature thus remains to be determined.

\begin{figure}
\psfig{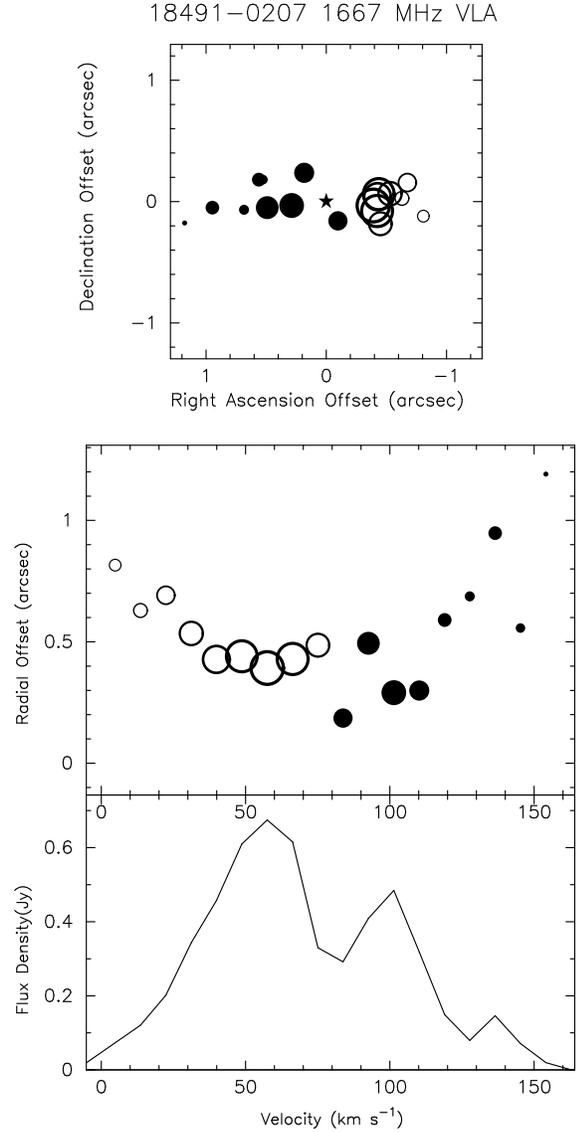}
\caption{IRAS 18491$-$0207:  $x$-$y$, $R$-$V$ and $I$-$V$ diagrams
for the OH 1612 MHz maser emission, from VLA observations taken in 1988.
}
\end{figure}

The OH 1667 MHz maser emission from IRAS 18491$-$0207 covers an
extreme width of 150 km s$^{-1}$. The VLA data (Fig.\ 12) are limited
by the very poor velocity resolution of 8.8 km s$^{-1}$. The stellar
position is assumed to coincide with the unweighted emission
centroid. The OH masers are unresolved in declination but show a
systematic velocity gradient of $\sim$ 75 km s$^{-1}$ arcsec $^{-1}$
in right ascension. An axi-symmetric or bipolar distribution is
indicated by the separation of the blue and red-shifted masers. We
interpret the maser distributions as a linear outflow source, with a
bipolar maser distribution similar to that of HD 101584 and IRAS
16342$-$3814. Higher velocity and spatial resolution maser
observations would be well worthwhile.

\subsection{IRAS 22036+5306} 

Oudmaijer et al. (1992) identified the IRAS source with HD 235718
(SAO34043). However, the present observations show the OH emission to
be located 40$^{\prime\prime}$ south of the star (see also Meixner et
al. 1999).  Coulson, Walther \&\ Dent (1998) report a small, very red
nebulosity near this position with near-infrared magnitudes of
(J,H,K,L)=(11.8,9.6,7.4,5.2). The POSS O- and E-plates show a faint,
red pointlike source at this position. The infrared photometry can be
fitted using two blackbody curves with temperatures of 1000 K and 130
K. Near-infrared spectra of this source taken in 1992 show CO
bandheads at 2.3$\mu$m-2.5$\mu$m and Brackett-gamma emission at 2.167
$\mu$m, both signatures of mass loss, and a 10-$\mu$m spectrum showed
broad deep absorption between 8 and 12 $\mu$m possibly due to
ice-covered silicates (Coulson, priv. comm.).

\begin{figure}
\psfig{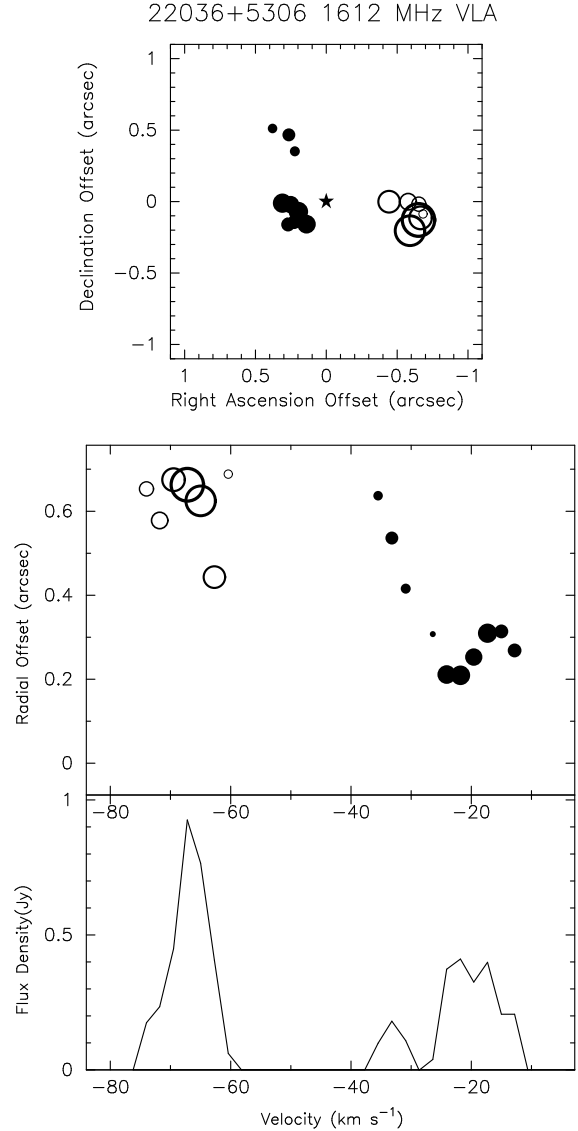}
\caption{(a) IRAS 22036+5306:  $x$-$y$, $R$-$V$ and $I$-$V$ diagrams
for the OH 1612 MHz maser emission, from VLA observations taken in
1990.
}
\end{figure}

\setcounter{figure}{12}
\begin{figure}
\psfig{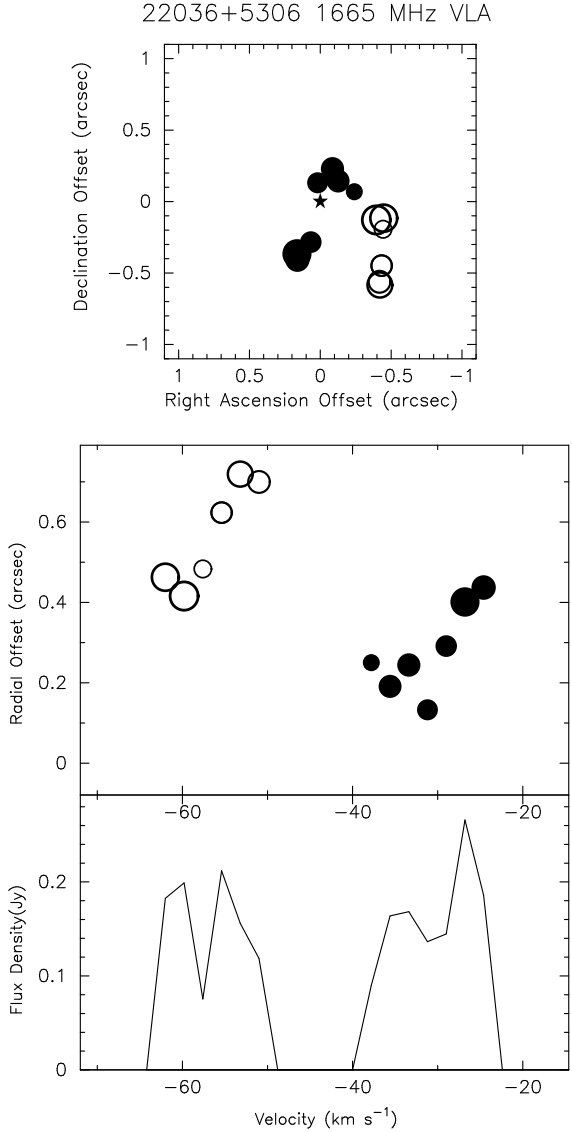}
\caption{(b) IRAS 22036+5306:  $x$-$y$, $R$-$V$ and $I$-$V$ diagrams for 
the OH 1665 MHz maser
emission, from VLA observations taken in 1990.
}
\end{figure}

The object is close to the Galactic plane and the central OH velocity
of $-43$ km s$^{-1}$ is consistent with a location in the Perseus arm.
The NVSS survey (Condon et al. 1998) did not reveal radio continuum
emission within $1^\prime$ of this position arguing against a classification
as an HII region. The characteristics of the OH masers and the absence of
a water maser (Zuckerman \&\ Lo 1987) are consistent with an evolved
object, perhaps similar to IRAS 16342$-$3814, but an evolved massive
star is also possible.
 
\setcounter{figure}{12}
\begin{figure}
\psfig{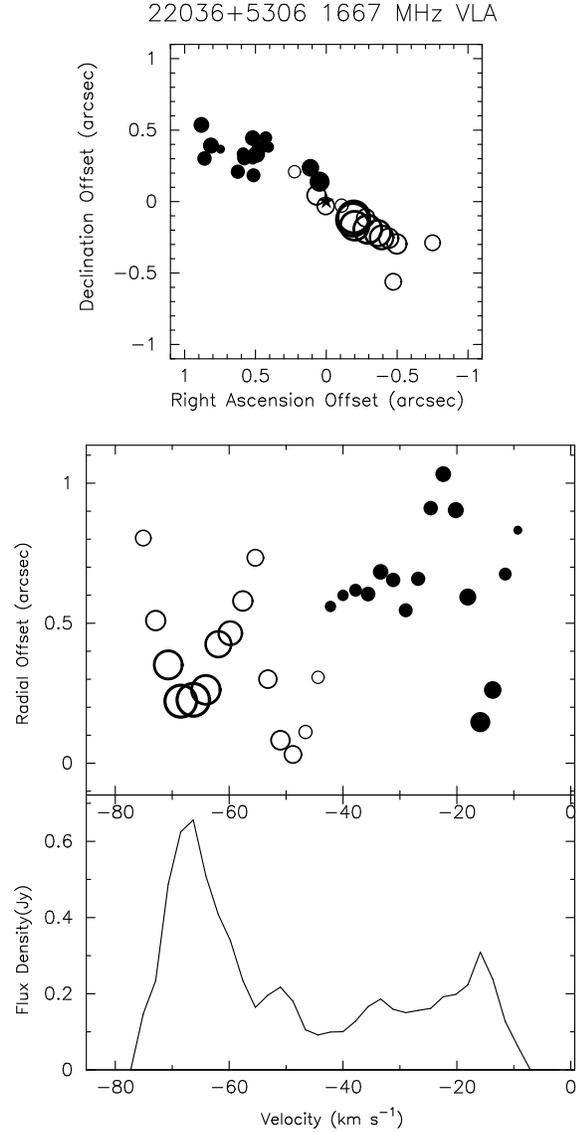}
\caption{(c) IRAS 22036+5306:  $x$-$y$, $R$-$V$ and $I$-$V$ diagrams for 
the OH 1667 MHz maser
emission, from VLA observations taken in 1990.
}
\end{figure}

The stellar position is not known and we have assumed
that it is at the unweighted emission centroid of all the detected
masers. From the centre of the 1667 MHz spectral profile, which
extends over the largest velocity range of $\sim$ 70 km s$^{-1}$, we
take the stellar velocity to be $-43\pm3$ km s$^{-1}$.

Figs 13a--c show axisymmetric distribution of the OH masers, with a
separation of the blue and red-shifted maser components present in all
three lines. The OH 1667 MHz masers are located within a linear
structure extending over 1.8 arcsec. The $R$-$V$ diagram for the 1667
MHz masers however is more complicated and indicates an incomplete
shell-like feature at velocities between $-60$ and $-25$ km s$^{-1}$,
while the masers at the extreme velocities appear to be more linear in
$R$-$V$.  The angular diameter of the OH emission is only marginally
consistent with the unresolved 10-$\mu$m counterpart (Meixner et al.\
1999).

The OH 1665 MHz and 1612 MHz masers appear to be located within a
small number of clumpy regions which may form part of the same
structures seen in the 1667 MHz line. Each of the three
transitions occupies different regions of the $x-y$ plane as expected
if the masers are located in clumpy regions in which different
physical conditions favour one of the OH transitions. The OH 1665 MHz
masers cover the smallest velocity range of $\sim$ 45 km s$^{-1}$ and
also appear to be located closest to the star. The 1612 MHz masers are
more extended in both space and velocity. Overall, we interpret the OH
masers of IRAS 22036+5306 as a combination of shell and linear outflow
but the separation between these components is not very clear.

IRAS 22036+5306 shows strong evolution in the spectral
profiles between the earlier data of te Lintel Hekkert (1988) and the
present data (1990). At both 1612 and 1667 MHz the low velocity
features, between --80 km s$^{-1}$ and --60 km s$^{-1}$, have
strengthened significantly while the high velocity maser feature,
between --25 km s$^{-1}$ and --10 km s$^{-1}$ has weakened. At the same
time, a plateau of emission has emerged.  In 1990, the OH 1667 MHz
spectral profiles of this source was very similar to the high outflow
sources OH231.8+4.2 and IRAS 15405$-$4945.  Unfortunately, no
polarization information is available.

\subsection{Summary: the two components}

The OH maser structures indicate that at least one of the two
components predicted by the model are present in the majority of the
sources.

Six objects show a shell/torus component. The expansion velocities for
this component listed in Table 3 show limited expansion velocities
mostly of 20--25 km s$^{-1}$.  OH231.8+4.2 and IRAS 08005$-$2356 have shell
expansion velocities which are too high for an AGB wind.  CO emission
from disks around evolved stars shows typical expansion velocities
around 7 km s$^{-1}$ (e.g. Olofsson \&\ Nyman 1999, Zweigle et
al. 1997).  This suggests that the OH torus may not be the original
circumstellar AGB torus but traces gas that has already been
accelerated.  For IRAS 16342$-$3814, the infrared model shows that the
density in its disk is sufficiently high to collisionally quench the
OH maser inversion.  This may also be the case for the other three
sources which show a linear outflow but not a shell or torus.

All but two objects (IRAS 17423$-$1755 and IRAS 08005$-$2356) show
evidence for a `linear outflow', with velocity increasing linearly
with distance from the star.  Extreme velocities with respect to the
star range from 15 to 80 km s$^{-1}$. Where both components are seen,
the linear outflow velocity reaches about twice the velocity of the
shell component. Deprojection may make the ratio a little larger.  The
observed velocity ratio between shell and linear outflow is in good
agreement with the predicted ratio of about a factor of 2 from the
interacting-wind model. The spatial size of the linear jet tends to be
smaller than the shell (except for OH231.8+4.2 where they are
similar).  The linear outflow appears to be seen close to the line of
sight, which suggests that the radial masers are seen and that
velocity deprojection effects are minor.  The linear outflow appears
to begin at the radius and velocity of the shell, in those cases where
both components are seen. This is in agreement with the geometry
assumed in section 4.2.

Strong circular polarization (25 per cent or more) of the OH maser
lines is seen in IRAS16342$-$3814, and in OH231.8+4.2, and five
sources (IRAS16342$-$3814, IRAS15405$-$4945, IRAS22036+5306,
OH231.8+4.2 (Morris \& Bowers 1980), and Roberts 22 (Allen et
al. 1980)) show variability in the ratio between different velocity
components. Such effects are not seen in normal AGB stars (e.g., Cohen
et al. 1987). The variability of upto a factor of 2 suggests the
effect may be related to line overlap between different Zeeman
components along the line of sight (Deguchi \&\ Watson 1986).

\section{Discussion}

\subsection{Nebulae}

The IRAS colour--colour diagram (Fig. 1) shows evidence for
two distinct groups, one with with colours typical for outflow sources
and the other with PostAGB colours.

All objects where imaging has shown a well-developed bipolar outflow,
are indeed found among the outflow sources. The optical morphology of
remaining two outflow sources (IRAS 22036+5306 and IRAS 15404$-$4945)
is not known. The outflow sources exhibit a narrow range in the
60/25-$\mu$m flux ratio. The 25/12-$\mu$m micron flux ratio varies by
a factor of 4: our modeling of IRAS 16342$-$3814 (Section 5.6.1)
suggests this reflects a range in the optical depth of the silicate
feature, which will depend on disk density and inclination.  HD 101584
may be related to this group but with a much lower disk optical depth.
Its CO emission indicates a similarly well-developed bipolar
structure.  The sources with the highest OH linear outflow velocities
are located at the red end of the outflow box; two of these lack a
shell component. If these sources have edge-on disks (as suggested by
the red colour), the high projected outflow velocity would indicate
that the outflow is located near the surface of the disk, close to the
equatorial plane, with a large opening angle.

The two sources with Post-AGB colours, IRAS 08005$-$2356 and Roberts
22, appear to exhibit bipolar outflows which are much less
collimated. A possible explanation could be that their circumstellar
disks are wider and less dense. Both objects show evidence for
chemical evolution and for the onset of central ionization.
Thus, these nebulae appear physically very different from
the outflow sources, although with a similar wind--wind interaction.

\subsection{Stars}

Spectral types are known for five stars in our sample (Table 3). The
star of IRAS 16342$-$3814 has not been detected but our infrared
modeling favours spectral class B, based on its non-detection.  The
central ionization indicated for IRAS 08005$-$2356 and Roberts 22
would require earlier types than listed, with temperature of 15000K or
a little higher.  This agrees with Bakker et al. (1997; see subsection
5.2 above). For temperatures above 18000 K extensive ionization is
expected which is not observed (e.g. Zijlstra et al. 1989). No
correlation is seen between derived temperatures and position in the
IRAS colour--colour diagram.

The observed wind velocities of the central stars, based on H$\beta$,
are 450 km s$^{-1}$ for Roberts 22 and $\sim 250 $ km s$^{-1}$ for He
3-1475 (from the spectra in Bobrowski et al.  1995) but only 130 km
s$^{-1}$ for HD 101584 (Bakker et al. 1996a) (much less than the
velocities of the molecular outflow). Such velocities are found in
post-AGB stars of spectral class late B, although the wind velocity
depends on the luminosity (Pauldrach et al. 1988) which is not known.
The early spectral types may be related to our selection criterium of
high OH outflow velocities which in our model requires a fast
stellar wind. In this case, most of the obscured, undetected stars in
our sample would require similar spectral types.

The Mira star of OH231.8+4.2 is the exception, and its fast wind is
likely to originate from a stellar companion which must have spectral
type of B or earlier. Cohen et al.  (1985) find some evidence for a
blue companion. The turn-off mass of its parent cluster excludes a
B-type main sequence star and so the companion would be a post-AGB star
itself. OH231.8+4.2, far from being the proto-type of its class, is in fact a
major exception. We suggest a classification as a symbiotic OH/IR
star.

Within our sample, only HD 101584 is an observationally confirmed
binary (Bakker et al. 1996a). Bipolar nebulae are often argued to be
related to binary evolution (e.g. Soker 1998), but direct evidence for
such a connection is missing.

\subsection{Evolution}

An important question is whether an evolutionary sequence can be
defined linking the bipolar OH/IR stars to PNe and to AGB stars.
We show that two such links may be present: the first with extreme
bipolar PNe, and the second with the IR-[WC] stars.

\subsubsection{Bipolar PNe}

Strong bipolar morphologies are very rare among PNe: most objects show
indication only for mild bipolarity or elliptical structures.  There
are six PNe known with morphologies resembling those of the outflow
sources (NGC6302, M2-9, Hb4, He2-437, 19W32 and K3-35; the
classification of K3-35 is uncertain). Their IRAS colours, indicated
by the triangles in Fig. 1, agree surprisingly well with the 'outflow'
sequence of the bipolar OH/IR stars. The similarity in morphology and
colours make a connection is likely, either by direct evolution (in
which case the star would have to evolve to higher temperatures on
short time scales compared to the colour evolution of the nebula), or
the PNe may be closely related objects but with faster evolving
central stars. The peculiar IRAS colours of the bipolar PNe has
previously been noted by Corradi \&\ Schwarz (1995).

Two of the PNe, NGC6302 and K3-35, show OH maser emission (only two
other PNe are known to show OH, one of which is discussed below).
They are the objects with the reddest 25/12 micron colours, and
therefore likely the highest optical depth through the disk.  The OH
strengthens the case for a connection between the outflow sources and
this rare group of PNe. Whether on longer time scales these objects
evolve into more familiar PNe is not known. There are more PNe known
with less distinct bipolar structures which could be a later
evolutionary phase, however a relation with the far more numerous
elliptical PNe is unlikely (e.g. Corradi \&\ Schwarz 1995).

\subsubsection{The IR-[WC] stars}

The chemical dichotomy seen in two of our objects is shared by a few
infrared-bright late [WC] stars (e.g. Zijlstra et al. 1991, Zijlstra
2000). The open squares in Fig. 1 show that these IR-[WC] stars have
IRAS colours closely similar to IRAS 08005$-$2356 and Roberts 22.  The
IR-[WC] stars have temperatures only slightly higher than the stars in
the present sample.  The possible ionized central region may indicate
that Roberts 22 is already becoming a planetary nebula.  One of the
IR-[WC] stars, IRAS 07027$-$7934, has a strong OH maser (Zijlstra et
al. 1991)

The combination of the dual chemistry, which is known for no other PN,
the closely similar stellar temperatures, and the OH emission of one
of the IR-[WC] stars, argues for a close relation between the IR-[WC]
stars, and Roberts 22 and IRAS 08005$-$2356. The main difference is
however that the [WC] stars have little or no hydrogen, while Roberts
22 shows strong hydrogen emission. To proof an evolutionary connection
would thus require a mechanism for removing the surface hydrogen, e.g.
mixing it down into the lower hydrogen-poor layers (e.g. Herwig 2000).

It is not evident that a dual-chemistry source can evolve into a
PN with a single (C or O-dominated) chemistry. Roberts 22 is the
best candidate at present for a [WC] star progenitor, but further
evolution into a typical PN seems unlikely.

We thus find two possible evolutionary sequences: one leading to the
few PNe with extreme bipolar morphologies, and a second leading to
the (equally rare) IR-[WC] stars. 

\subsubsection{Progenitors}

The only object known among the OH/IR stars which could be considered
a {\it progenitor} of the bipolar OH/IR stars, is OH 19.2$-$1.0.
Chapman (1988) modeled this star using a bipolar outflow model; the
MERLIN data indicates a velocity gradient across the source.  OH
19.2$-$1.0 is variable with a period of 600 days, and this and also
its IRAS colours (Fig. 6 in Chapman 1988) assign it to the tip of the
AGB. When this object will leave the AGB with declining mass loss, it
could become very similar to the objects studied in this sample. 
The outflow velocity of OH 19.2$-$1.0 is just below 50 km
s$^{-1}$. Thus, it is possible that our criterium of high outflow
velocity introduced a bias against AGB stars. 

Only a very small percentage of the OH/IR stars show irregular spectra
similar to OH19.2$-$1.0.  This would imply that well-developed bipolar
outflows are much more common in the post-AGB evolution than on the
AGB. 

Corradi (1995) makes a strong case for a relation between strongly
bipolar PNe and symbiotic stars. This possibility should also be
explored for the bipolar OH/IR stars. If its binarity is confirmed,
specially OH231.8+4.2 would be a candidate for a classification as a
symbotic OH/IR star.

The Red Rectangle also shows mixed chemistry around a Post-AGB star
(Waters et al. 1998a). It has a very old ($10^5$ yr) disk and no OH
emission, and is known to be a binary with a period of 318 days
(Waelkens et al. 1996). It is unlikely to be a progenitor of the
bipolar OH/IR stars, but may have a related origin.

\subsection{Post-AGB life times and evolution}

If the bipolar OH/IR stars are the main progenitors of PNe and white
dwarfs, their observed space density can be used to infer life times
of this phase.

The birth rate of white dwarfs (Iben \&\ Laughlin 1989) corresponds to
on average 1 object per hundred year within 2 kpc of the Sun. The rate
for PNe is the same or slightly higher for a local column density of
$35\pm10\,\rm kpc^{-2}$ and an observable lifetime of $5\times10^4\,$yr
(Zijlstra \&\ Pottasch 1991).

Within our sample, three objects are very likely to have distances
less than 2kpc: OH231.8+4.2 (1.3kpc), HD 101584 (0.6--1.1 kpc,
Hipparcos parallax) and IRAS 16342$-$3814 (0.7 kpc).  Roberts 22 is
likely to be about 2 kpc distant or less (Sahai et al. 1999a).  For
all four objects, the total in-band IRAS flux is incompatible with a
distance larger than 2 kpc, assuming post-AGB stellar
luminosities. The remaining objects in Table 2 are fainter, but there
are a few additional post-AGB objects in Table 1 (CRL 618, SAO 163075)
where the IRAS flux indicates distances less than 2 kpc. All are
within 200 pc of the Galactic plane and likely derive from the main
disk population.

The transition time from AGB to PN can be calculated:

$$ t_{\rm tr} = \frac{N}{f_{\rm p} f_{\rm o} }\times 100\,{\rm yr} \eqno (15), $$

\noindent where $N$ is the number of observed objects within 2 kpc of
the Sun, $f_{\rm p}$ is the fraction of the white dwarf population
passing through this phase and $f_{\rm o}$ is the fraction of the
transition time for which these objects have high-velocity OH
outflows.

Based on the objects above, we take $N=6$. We assume that $f_{\rm o}=0.5$,
since the derived velocities of the fast wind of the stars require
stellar temperatures above approximately 8000 K, as also indicated by
the observed stellar types.  During the first half of the transition
time this suggests that the linear outflow has velocities less than
our selection criterium requires.  This gives a minimum life time of
1200 yr. To estimate $f_{\rm p}$, we note that 14 per cent of PNe are 
bipolar and a further 4 per cent show point-symmetric structures
(Corradi \&\ Schwarz 1995) (the fraction of strongly bipolar PNe is
much less).  This indicates $f_p=0.2$ or less. We obtain
that the post-AGB transition time for the bipolar OH/IR stars is
$t_{\rm tr} \ge 7\times 10^3$ yr, with a probable value in excess of
$10^4$ yr. 

Such a large value is not compatible with the dense shells implied by
the IRAS colours (e.g. Siebenmorgen et al. 1994) or the expected $\sim
1000$ yr during which OH would remain observable. It should also be
compared with the dynamical age of the shells, using an inner radius
of $10^{16}$ cm (e.g. Table 4) and an expansion velocity of 7 km
s$^{-1}$ which yields about 500 yr. The dynamical age of the linear
outflows is even less. {\it We conclude that the bipolar OH/IR stars
remain in this evolutionary phase far longer than the dynamical age of
the envelope suggest.} A near-stationary reservoir of molecular gas,
such as a Keplerian disk, could act as a continuous source for the
observed structures. This would go beyond the model of Section
4, since non-radial flows of gas would be required to transport the OH
from a source in the disk to the bi-cones. Alternatively, observed
life times can be extended if phases of fast and slow winds alternate,
allowing the structures to re-form.

The main uncertainty in the time scales comes from $f_{\rm p}$, since
$N$ is based on observed numbers and cannot easily be lower than
assumed here.  If {\it all} post-AGB stars pass through the
bipolar/disk phase, $f_p=1$ and $t_{\rm tr} \approx 10^3$ yr.  But a
high value for $f_p$ is implausible, based on the distinct IRAS
colours in our sample, and the lack of indications for disks in the
majority of PNe. However, the derived time scale is an average and it
is not necessary to infer that all objects in our sample evolve so
slowly. If a minority of post-AGB stars show delayed evolution, they
could still dominate the observed numbers and increase the average
$t_{\rm tr}$. Thus, the post-AGB sample is likely a mixture of objects
with a range of transition times.

Based on Post-AGB evolutionary tracks, the stars should  evolve to
much higher temperatures within the long timescales. Slower evolution
of the star is possible if the core mass is very low (as suggested by
Bobrowski et al. 1995) rather than high (as suggested by Mellema
1997). A different, self-consistent model to explain the apparent
link between a slow star and a slow nebula is presented below.

\subsection{Post-AGB accretion}

The Red Rectangle is a known case where a post-AGB star is surrounded
by a near-stationary (circum-binary) disk. The surface abundances of
its star show that dust-depleted gas from its disk has been accreted
unto the star (Van Winckel et al. 1992, Waelkens et al. 1996). If the
present sample shows circumstellar disks similar to the Red Rectangle,
the possibility of accretion should be considered.

The increase in $T_{\rm eff}$ during the Post-AGB evolution is caused
by the slow decrease in stellar envelope mass due to nuclear burning
and Post-AGB mass loss.  Accretion will add gas to the envelope.  This
will have little effect on the AGB evolution, when the mass loss
dominates the evolution, but can can become significant during the
Post-AGB phase.

The Post-AGB mass-loss rate slowly decreases with increasing
temperature (Pauldrach et al. 1988, Bl\"ocker et al. 1995).  If at
some temperature the accretion rate equals the sum of the
nuclear-burning rate and the mass-loss rate, evolution to higher
temperature will cease.  The required accretion rate is of the order
of $10^{-7}\, \rm M_\odot \, yr^{-1}$. The subsequent evolution will
occur on the time scale of growth in core mass (causing the luminosity
to increase) rather than the time scale for envelope removal. This
would slow down post-AGB evolution by an order of magnitude.

Accretion could thus explain  Post-AGB transition times of
$\sim 10^4$ yr, but it requires the presence of non-expanding gas
near the star. Thus, accretion self-consistently links the slow nebula
with the evolutionary-challenged star.  Even if only a small number of
objects would experience accretion, the greatly increased lifetime
would mean that such objects could dominate in observed number
density.  This would yield a possible explanation for the much higher
occurrence of bipolar outflows among the post-AGB stars compared to
the AGB stars.

The effects of accretion could explain many of the observed
characteristics of the extreme bipolar objects, and link these
characteristics to the presence of a circumstellar disk.  The slow
disk itself remains unexplained. In the Red Rectangle and related
stars, it is believed to be formed due to the interaction between the
binary companions. This possibility should be investigated for the
present sample as well.

\section{Conclusions}

We have presented an investigation into the onset of bipolar structure
formation for OH/IR stars. The selected stars show irregular OH
spectra with high expansion velocities, consistent with a possible
classification as post-AGB star. We postulate an interaction between
an outer slow wind and an inner, somewhat faster wind, with at least
one component showing a dependency of mass-loss rate or expansion
velocity with polar angle.  The main conclusions are:

1. In the presence of a post-AGB--AGB wind--wind interaction, the
velocity of the swept-up shell is predicted to increase linearly with
distance from the star. The OH data confirms the presence of such a
linear outflow component.  A torus-like component is also seen with
expansion velocity of typically 25 km $s^{-1}$. 

2. All objects with a well-defined bipolar morphology and
high-velocity outflows have IRAS colours closely similar to those of
young outflow sources. This indicates that similar morphologies may be
present in both evolutionary phases.

3. Two of the selected stars show evidence for a chemical dichotomy, with
both crystalline silicates and PAHs or other carbon molecules being present.
The silicates are probably located in the torus (as shown by the OH data for
Roberts 22). 

4. IRAS 16342$-$3814 shows four point-like sources located symmetrically around
the central star, at a distance consistent with the outer edge of the
circumstellar torus. The nature of these (scattering?) sources is not
clear.

5. For five objects where data for the central stars are available,
the observed spectral types are M9, F5, A2, B9 and Be. However, the F star
has been suggested to have an earlier underlying type (late or early B),
affected by an extended atmosphere. The A star (Roberts 22) shows
indications for ionization also suggesting an earlier type. For IRAS
16342$-$3814, modeling implies a central star with a temperature
around 15 kK, or late B.  The present wind velocities of up to a few
100 km s$^{-1}$ appear consistent with spectral type around late B.

6. Two evolutionary sequences are indicated. The strongly bipolar
objects appear related to the (few) planetary nebulae with
similar morphologies and may constitute their progenitors. Two objects
show colours and chemistry similar to the infrared-bright, late
IR-[WC] stars and may be evolutionary related. One bipolar OH/IR star
is known on the AGB (OH 19.2$-$1.0), which is a possible progenitor
of the bipolar post-AGB stars.

7. OH231.8+4.2 has a Mira central star but also a fast wind. Its wind
velocity also indicates a B-star. The central star is therefore likely
a binary.  The presence of a Mira precludes a B-type main sequence
star, and therefore the companion appears to be a post-AGB
star. Following Cohen et al. 1985, we suggest OH231.8+4.2 is a
symbiotic OH/IR star.

8. The lifetime of this phase of bipolar post-AGB OH/IR stars is
found to be of the order of $10^4$ yr. This is much longer than
expected from normal post-AGB evolution. Accretion from a
circumstellar torus on a post-AGB star could stall its evolution near
the observed spectral types. Such a process could explain the apparent
link between retained disks and retarded stellar evolution.

\section*{Acknowledgments}
Some of the associations of the detected IRAS sources are based on the
SIMBAD data retrieval system, database of the Strasbourg, France,
astronomical Data Center.  The National Radio Astronomy Observatory is
operated under cooperative agreement with the National Science
Foundation.  The Australia Telescope National Facility is funded by
the Commonwealth of Australia for operation as a National Facility
managed by CSIRO.  MERLIN is operated by the University of Manchester
as a national facility of the Particle Physics \&\ Astronomy Research
Council.  IRAS data were obtained using the IRAS database server
(IRAS Software Telescope) of the Space Research Organisation of the
Netherlands (SRON), funded by the Netherlands Organisation for
Scientific Research (NWO) and also partly funded through the Air Force
Office of Scientific Research, grants AFOSR 86-0140 and AFOSR 89-0320.
This research has been supported by dedicated grants from the Leidse
Kerkhoven Bosscha Fonds, SAAO, MSSSO and ESO.  The VLA staff is
thanked for their assistance, and for their efficient archive.  We
have also made use of the HST and the La Palma archives.  Griet Van de
Steene is thanked for help with the reduction of the data.  Richard
Hook is acknowledged for helping with the astrometry of the HST image
of IRAS 16342$-$3814. Hans Olofsson is thanked for the use of the
CO data on HD 101584. We thank Jim Caswell and Miller Goss for useful
discussions and Lars-Ake Nyman for very helpful comments on this paper.
This paper has grown primarily out of the original work (and thesis) 
of Peter te Lintel Hekkert.

\end{document}

\newpage

{\bf Figure captions}

{\bf Figure 1} {IRAS colour--colour plot.  R21 = $\log{F_{25}/F_{12}}$
and R32 = $\log{F_{60}/F_{25}}$, with the IRAS fluxes not
colour-corrected. Regions for different types of objects are taken
from Pottasch et al.\ (1988), Zijlstra (1991) and Emerson (1987).
Filled circles are objects from Table 1. Objects discussed in this
paper are encircled. The open triangles show the extreme bipolar
planetary nebulae, NGC6302, M2-9, Hb4, He2-437, 19W32, K3-35. The open
squares show the [WC] stars with mixed C/O chemistry: BD+30$^{\rm
o}$3639, IRAS07027$-$7932, He2-113, CPD$-$56$^{\rm o}$8032 (Waters et
al.\ 1998b; Zijlstra et al.\ 1991; Cohen et al.\ 1998)}

{\bf Figure 2}{ Predicted velocity--radius relations for the model of
an hourglass-shaped bipolar outflow.\\ 
(a) The geometric model
consisting of an AGB shell and a biconal, symmetric
outflow. Velocities are proportional with distance from the star at
each point, and are radial with respect to the star.  The dotted lines
show the location of the strongest radially-beamed emission with the
highest observed velocity gradient; the dot-dashed lines show the opposite side
of each with lowest observed velocity gradient. The separate panel shows a
velocity-radius diagram with the arrows indicating the contributions
from the different locations in the model. The elliptical distribution of points 
corresponds to the AGB shell.
\\
(b) Predicted  velocity--radius relations for various values
of $\Omega$ (the opening angle of the cone) and $i$ (the inclination
angle between the equatorial plane and the line of sight). 
See discussion in the text.}

{\bf Figure 3} {OH 1667 MHz channel maps of OH231.8+4.2 obtained from
VLA observations taken in 1988}. The velocity of each image is
labelled in the top-right corner. The restoring beam is shown in the
last box.

{\bf Figure 4} {OH231.8+4.2: (a) Diagnostic $x$-$y$, $R$-$V$ and $I$-$V$
diagrams for OH231.8+4.2. In this and similar plots, the top panel
shows the relative position of each emission component plotted
relative to the unweighted emission centroid. The open and filled
symbols show blue-shifted and red-shifted emission components
respectively.  The stellar position is indicated by the filled star.
The middle panel shows the projected radial offsets of the maser
components from the stellar position, plotted against observed
velocity (with respect to the local standard of rest). The bottom
panel shows the total flux density, $I$, seen in the channel maps at
each velocity.\\ 
(b) Bottom panel: an optical image taken through an
H$\alpha$ filter, retrieved from the La Palma archive.  The axes
labelling is in pixels and the scale is 0.55 arcsec per pixel. North is 
at the top and East to the left. \\
Top panel: Overlay of the OH maser positions on the central region of the
H$\alpha$ image. The blue- and red-shifted masers are shown as open
circles and filled triangles respectively.}

{\bf Figure 5} {IRAS 08005-2356: (a) $x$-$y$, $R$-$V$ and $I$-$V$
diagrams for the OH 1612 MHz maser emission, from VLA
observations taken in 1990.\\ 
(b) $x$-$y$, $R$-$V$ and $I$-$V$
diagrams for the OH 1612 MHz maser emission, obtained from MERLIN
observations taken in 1988.\\ 
( c) Diagnostic $x$-$y$, $R$-$V$ and $I$-$V$ 
diagrams for the OH 1667 MHz maser emission, obtained from VLA
observations taken in 1990.}

{\bf Figure 6} {Roberts 22: (a) $x$-$y$, $R$-$V$ and $I$-$V$  diagrams for
the OH 1665 MHz maser emission, from ATCA observations in 1992.
observations taken in 1990. The stellar position is assumed to
coincide with strongest blue-shifted 1665 MHz feature at -26.8 km
s$^{-1}$. In Figs 5a--c, the plot origin is chosen to be the
unweighted emission centroid of all the detected maser positions.\\
(b) $x$-$y$, $R$-$V$ and $I$-$V$  diagrams for the OH 1667 MHz maser
distribution, from ATCA observations in 1992.\\  
(c) Composite $x$-$y$ and
$R$-$V$ diagrams showing the ATCA results in Figs 5a and 5b together
with higher resolution 1612 and 1665 MHz data obtained with the
PTI in 1986/7. The adopted stellar
position, corresponding to the position of the 1665 MHz emission peak
has coordinates of (0.14,0). The velocity ranges of the strong
torus-like structure and the northern feature described in the text
are labelled.\\ 
(d) The SWS spectrum of Roberts 22. In panel A the
complete SWS spectrum is shown, together with the spline fit
continuum. In panel B the continuum-subtracted spectrum up to 15$\mu$m
is shown with the PAH-features indicated.  Panel C shows the
continuum-subtracted spectrum longwards of 15 $\mu$m; indicated are
the positions of crystalline silicate features.  The spectra were
reduced using the SWS off-line processing software, version 7.0.
Fringes in the 16.--29.2$\mu$m part of the spectrum were removed using
the InterActive (IA) data reduction package routine {\em
fringes}. Flux and wavelength calibration procedures are described in
Schaeidt et al. (1996) and Valentijn et al. (1996).}

{\bf Figure 7} {HD 101584: (a) $x$-$y$, $R$-$V$ and $I$-$V$ diagrams for the
OH 1667 MHz maser emission, obtained from ATCA observations taken in
1991. \\ (b) ATCA OH 1667 MHz maser positions plotted together with the
$^{12}$CO$J=2\rightarrow1$ positions from SEST observations in 1997
(Olofsson \& Nyman 1999). The velocity at each CO position is
labelled. The velocity ranges for the two groups of OH masers are
given in brackets. The stellar position is assumed to be at the
centroid of both the OH and CO distributions as indicated by the
filled star at the origin.}

{\bf Figure 8} {IRAS 15405$-$4945: (a) $x$-$y$, $R$-$V$ and $I$-$V$ diagrams
for the OH 1665 MHz maser emission, from ATCA observations taken in
1991. (b)  $x$-$y$, $R$-$V$ and $I$-$V$ diagrams for the OH 1667 MHz maser
emission, from ATCA observations taken in 1991.}

{\bf Figure 9} {IRAS 16342$-$3814: (a)  $x$-$y$, $R$-$V$ and $I$-$V$ diagrams
for the OH 1612 MHz maser emission, from VLA observations taken in
1990.\\ 
(b) An HST image through the V-filter (F555) taken with the
Planetary Camera in 1997, calibrated using the HST pipeline
software. North is up and east is to the left. The pixel scale is
0.045 arcsec.  The image has been de-rotated and corrected for field
distortions. The contour levels are in steps which increase by a
factor of 2.  The triangles indicate the positions of the OH
components. The filled star shows the estimated position of the
central star, at $\rm \alpha=16^h34^m17.06^s$, $\rm
\delta=-34^o18^\prime17.9^{\prime\prime}$ (B1950).  The error bars in
the corner indicate the uncertainty in alignment of radio and optical
coordinates.\\
(c) The infrared spectrum of IRAS 16342$-$3814, showing
the IRAS LRS spectrum, the IRAS photometry (filled circles), and
optical and infrared photometry (other symbols) of van der Veen et
al. (1988). The optical photometry appears to be wrong, possible
related to a nearby field star.  The fit consisting of star+disk+shell
is indicated, with the shell also shown separately.}

{\bf Figure 10} {IRAS 17253$-$2831:  $x$-$y$, $R$-$V$ and $I$-$V$ diagrams
for the OH 1612 MHz maser emission, from VLA observations taken in 1988.}

{\bf Figure 11} {IRAS 17423$-$1755:  $x$-$y$, $R$-$V$ and $I$-$V$ diagrams
for the OH 1667 MHz maser emission, from VLA observations taken in 1988.}

{\bf Figure 12} {IRAS 18491$-$0207:  $x$-$y$, $R$-$V$ and $I$-$V$ diagrams
for the OH 1612 MHz maser emission, from VLA observations taken in 1988.}

{\bf Figure 13} {IRAS 22036+5306: (a)  $x$-$y$, $R$-$V$ and $I$-$V$ diagrams
for the OH 1612 MHz maser emission, from VLA observations taken in
1990.\\ 
(b)  $x$-$y$, $R$-$V$ and $I$-$V$ diagrams for the OH 1665 MHz maser
emission, from VLA observations taken in 1990.\\ 
(c) $x$-$y$, $R$-$V$ and $I$-$V$ diagrams for the OH 1667 MHz maser emission, 
from VLA observations taken in 1990.}

\end{document}